\documentclass[sigconf]{acmart}
\usepackage{listings}
\usepackage{enumerate}
\usepackage{graphicx}
\usepackage{multirow}
\usepackage{array}
\newcommand{\m}{}
\patchcmd{\quote}
  {\list{}{\rightmargin\leftmargin}}
  {\list{}{\rightmargin1em \leftmargin1em}}
  {}{}

\setcopyright{cc}
\setcctype{by}
\acmJournal{IMWUT}
\acmYear{2025} \acmVolume{9} \acmNumber{3} \acmArticle{107} \acmMonth{9} \acmDOI{10.1145/3749484}
\acmConference[IMWUT '25]{Proceedings of the ACM on Interactive, Mobile, Wearable and Ubiquitous Technologies}{Vol. 9, No. 3, Article 107}{September 2025}
\acmISBN{}


\begin{document}
\title[HandProxy]{HandProxy: Expanding the Affordances of Speech Interfaces in Immersive Environments with a Virtual Proxy Hand}

\author{Chen Liang}
\email{clumich@umich.edu}
\affiliation{%
 \institution{University of Michigan}
 \city{Ann Arbor, MI}
 \country{USA}
}
\author{Yuxuan Liu}
\email{liurick@umich.edu}
\affiliation{%
 \institution{University of Michigan}
 \city{Ann Arbor, MI}
 \country{USA}
}
\author{Martez Mott}
\email{mamott@microsoft.com}
\affiliation{%
 \institution{Microsoft Research}
 \city{Redmond, WA}
 \country{USA}
}
\author{Anhong Guo}
\email{anhong@umich.edu}
\affiliation{%
 \institution{University of Michigan}
 \city{Ann Arbor, MI}
 \country{USA}
}

\begin{abstract}
Hand interactions are increasingly used as the primary input modality in immersive environments, but they are not always feasible due to situational impairments, motor limitations, and environmental constraints. Speech interfaces have been explored as an alternative to hand input in research and commercial solutions, but are limited to initiating basic hand gestures and system controls. We introduce HandProxy, a system that expands the affordances of speech interfaces to support expressive hand interactions. Instead of relying on predefined speech commands directly mapped to possible interactions, HandProxy enables users to control the movement of a virtual hand as an interaction proxy, allowing them to describe the intended interactions naturally while the system translates speech into a sequence of hand controls for real-time execution. A user study with 20 participants demonstrated that HandProxy effectively enabled diverse hand interactions in virtual environments, {\m achieving a 100\% task completion rate with an average of 1.09 attempts per speech command and 91.8\% command execution accuracy}, while supporting flexible, natural speech input with varying levels of control and granularity.
\end{abstract}

\settopmatter{printfolios=true}

\begin{teaserfigure}
  \includegraphics[width=\textwidth]{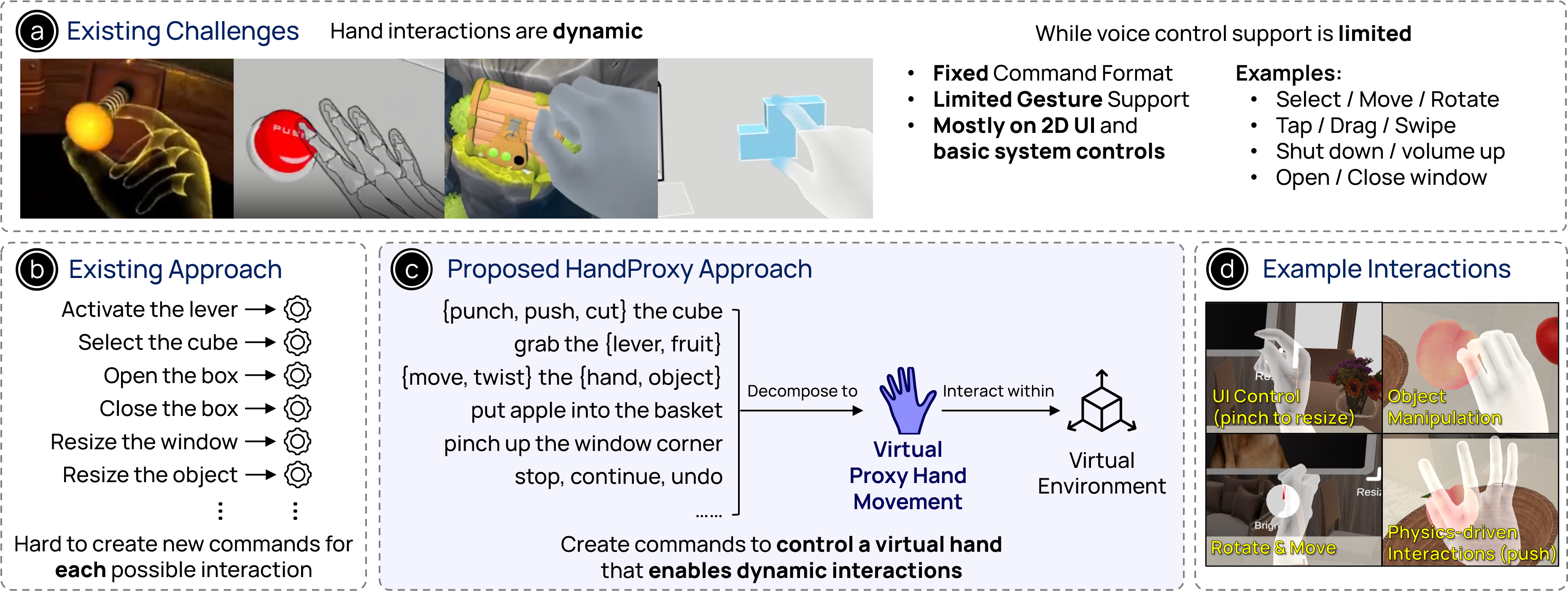}
  \caption{While XR devices increasingly support dynamic hand interactions, the corresponding speech-based interfaces remain limited (a). {\m While adding speech commands enhances interface capability (b), they lack scalability for diverse hand interactions. We introduce the approach of using a virtual proxy hand that enables users to describe their desired interactions, which the system translates into executable hand movements (c).} This approach supports various interactions, such as UI control, object manipulation, and interactions that emerge from physics,  collisions, or hand rotation and movement (d).}
  \Description{A figure showing the existing challenges, existing approach, proposed approach, and example interactions. (a) Existing Challenges: XR hand interactions are dynamic, including pulling a rod, pressing the button, opening a chest, and pinching a cube. However, current systems rely on fixed command formats, limited gesture support, and mostly 2D UI-based controls and system controls. Examples include selecting, moving, rotating, tapping, dragging, volume up/down, and closing UI windows. (b) Existing approaches mostly focusing on supporting additional functions through creating new command, but could get challenging to do so for each possible interaction. (c) We propose the HandProxy approach, where users can use speech commands to control a virtual hand, which then enables dynamic interactions. Users can describe interactions using natural language (e.g., "punch the cube," "grab the lever," "move the hand"), which are translated into virtual proxy hand movements. (d) With this approach, the HandProxy can be used for UI control (pinch to resize), object manipulation, rotation and movement, and physics-driven interactions (push). }
  \label{fig:teaser}
\end{teaserfigure}

\maketitle

\begin{figure*}[]
  \includegraphics[width=\linewidth]{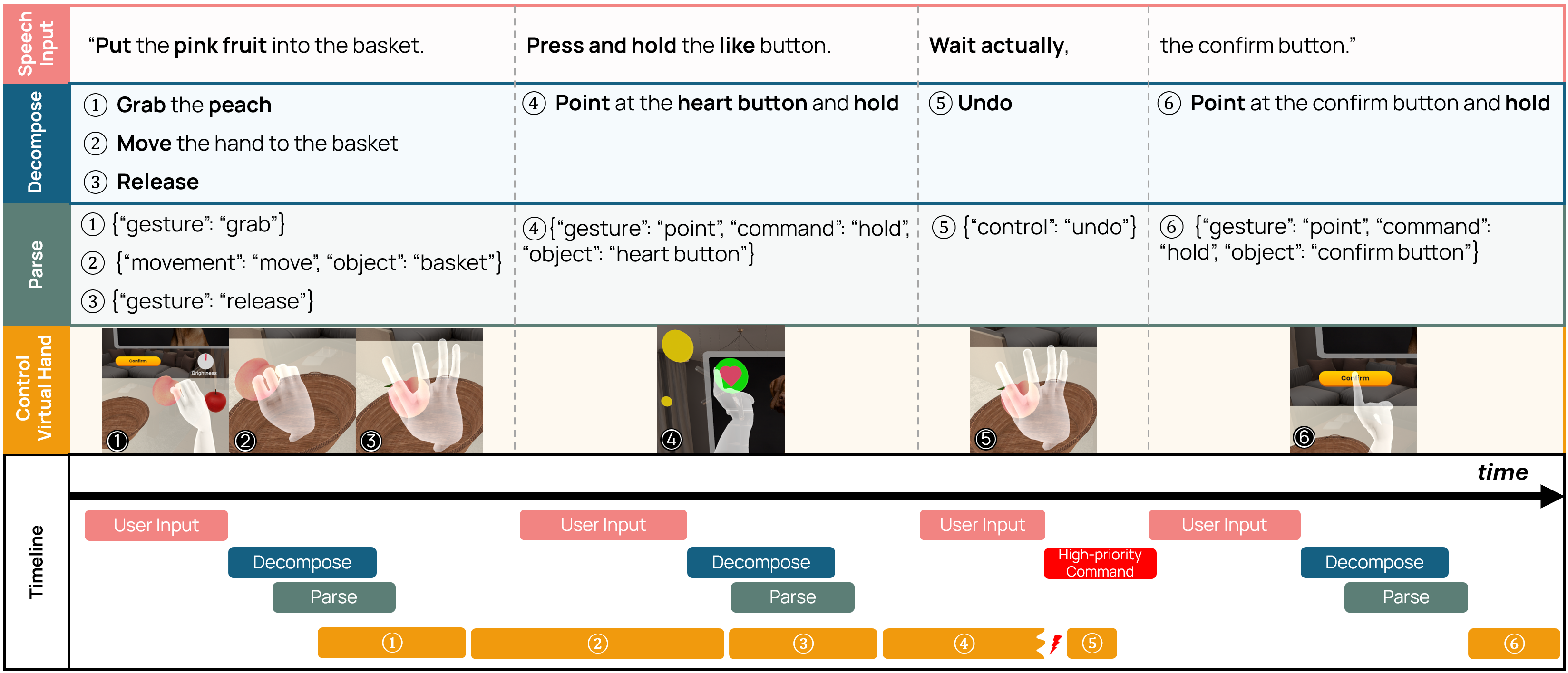}
  \vspace{-1.5pc}
  \caption{{\m Example of how HandProxy would process the user's speech commands. The user continuously talks to the system, where the system concurrently recognizes the commands, decomposes them into steps, parses them into executable hand control instructions, and calculates the sequence of hand movements to control the virtual hand performing the desired interaction. The processed instructions are added to a queue for continuous hand control. HandProxy utilizes both rule-based methods to detect high-priority commands (e.g., stop, undo) and large language model to interpret natural language input. Whenever a high-priority command is detected (e.g., (5) in the plot), HandProxy will terminate the current execution and perform the high-priority commands.}}
  \Description{A timeline showing how a user's speech input is processed. The user continuously saying the command "put the pink fruit into the basket, press and hold the like button, wait actually, the confirm button". The system splits it into 4 commands, decompose it into steps, and parses it into system executable instructions. For example, put the pink fruit into the basket is decomposed into (a) grab the peach, (b) move the hand to the basket, (c) release. Press and hold the like button is decomposed into one step of point at the heart button and hold. Wait actually is decomposed as undo. And "the confirm button" is decomposed as point at the confirm button and hold. All commands are processed continuously. When there is a high-priority command such as undo, it will break the current queue and execute it immediately. }
  \label{fig:intro_example}
\end{figure*}

\section{Introduction}
Hand tracking is increasingly used as the primary input modality on extended reality (XR) devices to interact with the virtual environment \cite{quest_gestures,hololens_gestures,apple_gestures}. As a result, many applications are built specifically for hand interactions, including games, productivity apps, and 2D/3D user interfaces (UI). For example, users can pinch and drag the corner of a window to resize UI windows on Meta Quest, or grab and move a virtual object to place it in an desired position in mixed reality on Apple Vision Pro. Games such as Waltz of the Wizard\footnote{\url{https://www.aldin.io/}} rely on various hand interactions, including grabbing and moving game objects, punching skulls, knocking doors, shaking hands, and tilting cups to pour water. These examples highlight the expressiveness and complexity of hand interactions beyond traditional input modalities, and demonstrate diverse interaction possibilities that could only be achieved with hands in the virtual environments.

However, users may not always be able to perform the expected hand interactions, making it challenging or even impossible to effectively interact with the virtual environments. This could be caused by situational impairments \cite{siid_wobbrock}, e.g., when users' hands are occupied with other physical tasks, they may not be able to manipulate virtual objects and interfaces. Environment constraints could also cause challenges, e.g., users may not be able to move freely in a confined physical space \cite{handycast}, and virtual objects could be out of physical reach in an enlarged virtual space \cite{ar_reach}. Additionally, predefined hand interactions may be inaccessible to users with upper body limitations \cite{ubl_gesture_customize, vr_ubl}. These challenges indicate the need for alternative ways of interacting with virtual environments, specifically those that can reproduce or achieve similar interaction experiences and the expressiveness as the original hand interactions. 

Among the various alternatives and enhancements to hand input, speech has emerged as an ideal modality due to its intuitiveness and high degree of freedom. Prior research has explored its uses in object manipulation \cite{llm_obj_mani}, navigation \cite{speech_nav}, and environment creation \cite{how_prompt_dis}. Additionally, speech has been adopted as an assistive input modality in many mainstream XR devices. For example, Apple Vision Pro provides voice control \cite{apple_voice_control} as an alternative to hand input, which can initiate simple interactions, including tap, swipe, and drag and drop on 2D interfaces. Users can also issue high-level commands, such as ``turn up the volume'', instead of doing it step-by-step. However, these speech interfaces typically support a limited set of hand interactions and require users to use a rigid, predefined command format to trigger actions. While this approach works for basic system-level interactions, it falls short in supporting the complex and diverse interactions a hand can do in virtual environments. {\m This limitation is particularly relevant to interactions triggered by hand gestures or movements, such as 3D object manipulation (e.g., pinch, punch, squeeze) and physics-based interactions (e.g., slice a fruit in half with a cutting gesture).} Thus, we seek to address the question: \textit{``How can we expand the affordances of speech interfaces to support expressive and diverse interactions, especially those that can be achieved via hand interactions in virtual environments?''}

Instead of merely designing more speech commands, this work proposes an alternative approach: \textbf{enabling users to use natural language to control a \textit{virtual hand}, where it can then simulate corresponding movements and perform necessary interactions on the user’s behalf.} This approach is motivated by the observation that many interactions result from a sequence of hand movements, where each step contributes specific meaning to the overall interaction. For example, when pulling a lever, visual effects are triggered as the hand grabs the handle, the lever moves as the hand moves, and the effect finally takes place when the hand releases it. Furthermore, many interactions are not explicitly defined but emerge as a result of other factors, such as collision and physics. For example, an application may not have a specific ``clean the table'' interaction defined, but the user can achieve this by sweeping the virtual hand across the table, pushing objects away through simulated collision. Therefore, it is impossible to simply create predefined speech commands for all possible interactions, as interactions themselves could be loosely defined, i.e., they are either context-dependent or dynamically generated. These considerations motivate our design choice of preserving the virtual hand within the environment. This allows users to control and reproduce actions as if real hand input is used, enabling them to perform interactions that mirror the capabilities of a physical hand.

To achieve this, we introduce HandProxy, a system that enables users to control a virtual hand through continuous, natural speech input. HandProxy is inspired by the concept of interaction proxy \cite{interaction-proxy}, where a virtual hand is used as the proxy layer between the speech interface and the immersive environment, as shown in \autoref{fig:teaser}. We explored existing hand interactions in prior work, commercial devices, and VR applications \cite{2d_3d_speech, how_prompt_dis, hands_free_VR}, and synthesized a list of hand control primitives that could be used for decomposing and reproducing common hand interactions. Within the comprehensive space of possible hand interactions, this work starts by focusing on the fundamental, while critical, use case --- one-handed interactions. Specifically, we investigate how one-handed manipulative interactions can be initiated through the speech interface and categorize them into four key primitives: gesture, target, spatial, and temporal control. These primitives are used as the foundation to reproduce various interaction, including both detailed controls (e.g., do a pinch gesture, grab the apple) or high-level interactions (e.g., maximize the volume). As shown in \autoref{fig:intro_example}, the system captures users’ natural speech, parses it into a list of executable commands with a Large Language Model (LLM), calculates the desired hand skeleton data, and renders it in the target system and application. HandProxy is optimized for real-time interaction, and we demonstrated that it can be used in a variety of interaction scenarios.

To understand how well HandProxy could perform hand manipulations from diverse natural language commands, and to gain insights from users' experiences of controlling a virtual hand through speech, we conducted a user study with 20 participants. Our study shows that HandProxy enabled users to complete a variety of hand interactions in the virtual environment, including different types of interactions such as mid-air gestures (e.g., swipe left), direct object manipulation (e.g., twist the knob), high-level interaction tasks (e.g., increase the volume), and with varying levels of complexity (e.g., one-step to multi-step interactions). {\m Participants reached a 100\% task completion rate, and took an average of 1.09 attempts for their speech commands to be correctly executed by the system (with a command execution accuracy of 91.8\%). HandProxy was able to handle diverse variations of participants' commands for the same tasks, such as using descriptive commands (e.g., ``touch your index finger and thumb'' to pinch, ``grab the red fruit'' to grab an apple), varying levels of details (e.g., to increase the volume, either grab the slider and move up, or directly say ``maximize the volume''), or sentence structures.} Furthermore, participants reported the system to be intuitive, effective, and require minimal learning to use, and pointed out possible improvements including more detailed feedback for enhanced disambiguation and discoverability, greater responsiveness, and supporting additional hand controls.

The specific contributions of our work therefore include:
\begin{enumerate}
    \item A set of primitives to categorize common hand interactions in the virtual environment, allowing hand interactions to be decomposed and reproduced.
    \item A real-time system, HandProxy, that enables users to issue natural speech commands to control a virtual hand to simulate and perform hand-based interactions in the virtual environment. 
    \item An investigation into the effectiveness and user experience of using speech to control hand movement for various interaction tasks in the virtual environment.
\end{enumerate}

\section{Related Work} 
Our work is based on the literature of hand gesture input in XR, alternative input modalities for hand interactions, speech interfaces, and interaction proxies. 

\subsection{Enhancements and Alternatives for Hand Interactions}
Hand interactions have been widely used as an input method for interacting with virtual environments as it is intuitive, expressive, and preferred by users \cite{gesture_natural_ar, gesture_preferred_game}. These advantages have made them a core interaction technique across various devices and application domains, including object manipulation \cite{obj_mani_tech, hand_spatial, grasp_shell, portal_ble}, virtual collaboration \cite{be_there}, creativity support \cite{magicalhands, hand-avatar, mixfab}, gaming \cite{meta_hand_apps}, and object retrieval \cite{hand-interface}. 

Despite these advantages, hand interactions also come with certain limitations. Situational impairments \cite{siid_wobbrock} such as occupied hands in AR, physical disabilities \cite{two_in_one, ubl_gesture_customize, older-adult-gesture} such as upper body limitations, fatigue caused by long-term use of mid-air gestures \cite{arm-fatigue, arm-fatigue-2}, or environment constraints such as the confined physical space \cite{handycast} or out-of-reach virtual objects \cite{ar_reach} could limit users' abilities to perform hand input. Additionally, users may prefer different gesture \cite{ar_ges_elicit_hologram, user-defined, ges_vocab_iter_haijun}, making the standardized gesture input less desirable in certain situations. 

To address these challenges, researchers have explored various enhancements and alternative interaction techniques. Some approaches expand hand input capabilities, such as extending the virtual hand for reaching distant objects \cite{gogo, ar_reach}, using multiple virtual hand copies for easier object selection \cite{ninja_hand}, and exploring customized gesture sets for users with motor impairments \cite{mingming_gesture}. Others focus on alternative input modalities, including sensor-based modalities \cite{ubl_gesture_customize, biosignal}, speech-based commands \cite{ve_speech}, and multi-modal interaction techniques, such as gaze + voice \cite{gaze_situational}, hand + gaze \cite{hand-gaze}, and hand + speech \cite{hand-speech}. However, they are either designed to address a limited scope of specific scenarios (e.g., basic object manipulation, locomotion) or may require complex input setups, which are less practical to be deployed on existing devices for a broad range of interaction scenarios. 

Building on insights from prior research, HandProxy uses speech as an alternative modality. This choice is motivated by the fact that speech is widely accessible across various devices and is already integrated as a built-in control mechanism in many commercial systems \cite{apple_voice_control, hololens_voice_control}. We explore how speech can be used to initiate an expressive and diverse range of interactions that usually require hand input. Rather than simply substituting individual gestures for speech commands, our work expands the capabilities of the existing speech interface through a proxy hand, enabling users to perform hand-based tasks through flexible and intuitive spoken input.

\subsection{Speech Interfaces in Immersive Environments}
{\m Speech has been widely adopted as a control mechanism across various scenarios. For example, users can issue supported voice commands to interact with mobile devices \cite{voicify_ui, voice_ios} or desktop user interfaces \cite{voice_as_sound, voice_mac}. In robotics, speech is used to control robot actions \cite{voicebot, hierarchy_speech_robot}. More recently, the integration of speech interfaces with large language models (LLMs) has expanded these capabilities \cite{chatgpt_robot, voicepilot, saycan, waypoint_llm, llm_motion_planning, llm_puppeteer, rehab_llm_robot}, enabling functions such as motion planning, task decomposition, and flexible natural language input.}

In immersive environments, speech interfaces have been shown to be particularly intuitive, expressive, and natural due to their high degree of freedom \cite{2d_3d_speech, ve_speech, voice_control_vr}. As a result, it has emerged as a promising input modality for interactions in virtual environments. Prior work has explored speech-based interactions for various tasks, including virtual objects manipulation \cite{voice_control_vr, put_that_there, speech_n_ges_obj_mani, hands_free_VR}, locomotion \cite{speech_nav}, and scene creation \cite{vr_copilot, how_prompt_dis}, using verbal or non-verbal commands such as breathing or sound actions \cite{breathvr, blowclick, apple_sound_action}. Commercial XR devices have also integrated speech as a built-in assistive input modality. For example, Apple Vision Pro \cite{apple_voice_control}, Meta Quest \cite{quest_speech_control}, and Microsoft HoloLens \cite{hololens_voice_control} allow users to issue system commands (e.g., volume adjustment, power control) or perform simple hand gestures (e.g., tap, swipe, drag-drop) using speech. While these approaches demonstrate the versatility of speech interfaces, they require users to follow rigid command structures, and the interactions supported are limited to basic hand gestures and system functions. 

{\m Recently, integrating large language models (LLMs) into speech interfaces has opened up diverse interaction possibilities, allowing users to directly create, manipulate, query, and engage with their environments through flexible speech input \cite{llmr, llm_obj_mani, how_prompt_dis}. Building upon these prior works, we specifically explore ways to expand the affordance of speech interfaces through a generalizable approach, so that it can enable more interaction possibilities, while requiring minimal direct modification of individual applications to support it. To achieve this, we introduce a virtual proxy hand in the pipeline.} By describing the movement of a proxy virtual hand in flexible speech input, we seek to enable expressive interactions in virtual environments, especially those that hands or only hands can do. Additionally, to design an effective speech-based system, we built upon prior work on speech interfaces \cite{vui_guidelines, voicepilot, 2d_3d_speech}, user expectations for AI-driven agents in VR \cite{how_prompt_dis}, and contextual considerations in speech systems \cite{context_aware_speech}. These insights inspire the design choices of HandProxy to incorporate embodied knowledge, contextual understanding, conversational memory, interaction state control, and common knowledge integration to enhance the system’s ability to interpret and execute user commands effectively. Combining these design choices, our system can effectively support users to describe their interaction intentions, and the proxy hand can then perform the necessary interactions on the user's behalf.

\subsection{Interaction Proxies}
Interaction proxies are the extra layer inserted between the original and the manifest interface to add or modify interactions without changing the app's source code \cite{interaction-proxy}. Their ability to introduce new functionality with minimal modifications makes them particularly valuable for tasks such as input remapping \cite{interaction-proxy, brushlens, interactiles, interactout} and UI automation \cite{sugilite}. This concept has also been applied to immersive environments. For example, researchers have created tangible proxies that map physical input with digital interactions \cite{ubitouch, annexing_reality, magicbook, holotouch, augment_everyday_obj}. {\m Other work explores remapping of complex 3D input motions in VR to a more accessible ranges of motion or simpler input devices \cite{motion_blocks}.} Some work specifically explored the remapping of hand interactions through the proxied interface in virtual environments, such as remapping VR hand interactions to mobile devices \cite{handycast} or finger movement \cite{fingermapper} to facilitate interactions in constrained spaces. Additionally, McGlashan et al. \cite{ve_speech} introduced proxy agents as an interaction metaphor in VR, allowing users to issue commands to virtual agents that execute tasks on their behalf.

Inspired by these ideas, HandProxy explores how expressive hand interactions can be effectively translated into speech commands. Rather than relying on a rigid set of predefined voice commands for each interaction, HandProxy allows users to control a virtual hand as a proxy, enabling it to perform actions just as a real hand would. This approach ensures that users can issue commands in a natural and intuitive way, while the virtual hand continues to interact with applications as expected. Also, it reduces the need for system-specific modifications to support the mapping and functionality of the speech interface, making this approach compatible and generalizable across different applications that accept hand input. 
\section{Primitives of Hand Interaction}
\label{par:building_blocks}
To effectively perform interactions using the virtual proxy hand, it is essential to define a control framework that can reproduce a wide range of hand movements. Anatomically, the human hand has 27 degrees of freedom (DoF), covering finger extension, flexion, abduction, adduction, wrist rotation and translation \cite{hand_anatomy}. Given this complexity, directly replicating all possible hand movements would be impractical for virtual interactions, as each of the 27 DoF needs to be mapped to an input for control. Therefore, in this section, we introduce a simplified control framework designed to balance usability and expressiveness, and ensure that commonly used hand interactions in virtual environments can be effectively reproduced while maintaining intuitive control.

Our framework is inspired by the concept of hierarchical gestures \cite{hierarchical_gesture_1, hierarchical_gesture_2}, where complex hand interactions can be formed by strategically combining multiple primitive gestures. However, in this work, we take a slightly different approach --- instead of building up interactions from simpler gestures, we explore how commonly used hand interactions can be decomposed into a small set of shared fundamental control primitives. The goal of this decomposition is to create a compact yet expressive control framework that simplifies the reconstruction of hand interactions while still accommodating a wide range of interaction possibilities. Additionally, this approach could align more naturally with speech input, which is inherently structured as a combination of smaller linguistic units (i.e., words). For example, given the command ``pinch up'', it can naturally be decomposed into a gesture of pinch and a movement of up, matching the primitive hand controls. With these characteristics, the framework simplifies the integration of hand control with speech input while also making interactions intuitive for users.

\begin{figure*}[]
  \includegraphics[width=.98\linewidth]{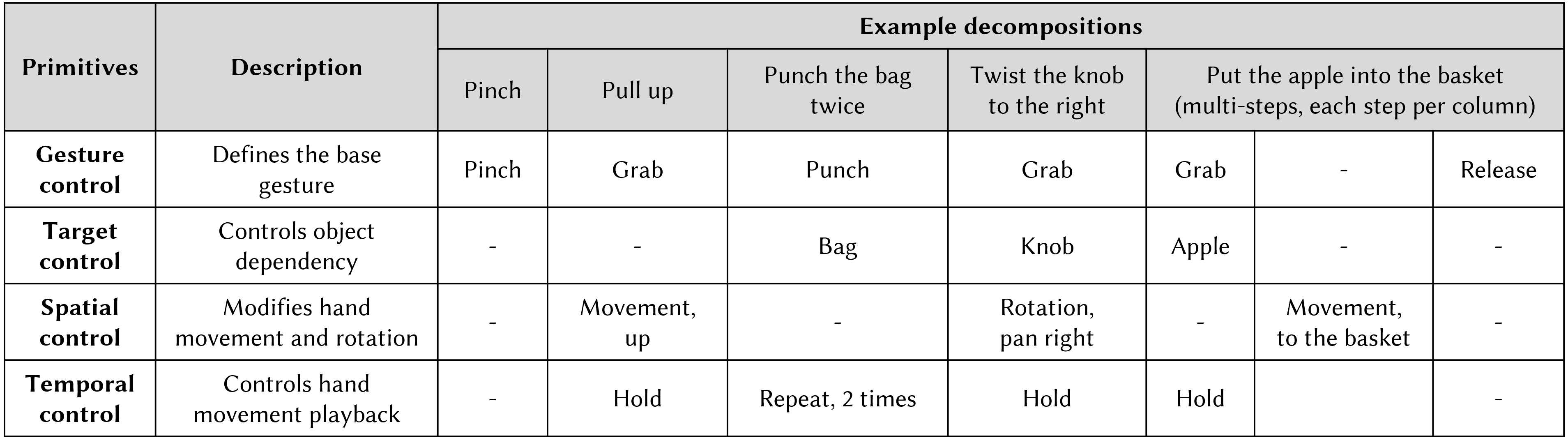}
  \vspace{-.5pc}
  \caption{{\m Examples of how hand interactions can be decomposed into one or multiple combinations of hand control primitives.}}
  \Description{A table showing how some commands can be decomposed using hand control primitives. For example, pinch can be decomposed into a gesture control of pinch. Pull up can be decomposed into a gesture control of grab, a spatial control of move up, and temporal control of hold. Punch the bag twice can be decomposed into a gesture control of punch, a target control of bag, and a temporal control of repeat 2 times. Twist the knob to the right can be decomposed into gesture control of grab, target control of knob, spatial control of rotate right, and temporal control of hold. Put the apple into the basket can be decomposed into three steps. First step has a gesture control of grab, a target control of apple, and a temporal control of hold. Second step has a spatial control of move to the basket. Third step has a gesture control of release.}
  \label{fig:primitives}
\end{figure*}

To develop this framework, we reference prior work on hand gesture definitions \cite{karam_gesture_taxonomy}, mid-air gestures \cite{user-defined, ar_ges_elicit_hologram, midair_ges_consensus}, gesture design considerations \cite{ges_vocab_iter_haijun}, as well as gesture vocabularies in XR device control \cite{apple_gestures, quest_gestures, hololens_gestures} and XR applications \cite{meta_hand_apps}. In this work, we focus on the fundamental and critical use case — one-handed manipulative interaction, and identified a set of fundamental hand control ``building blocks'' that served as the core components for reproducing diverse interactions, including \textit{gesture control}, \textit{target control}, \textit{spatial control}, and \textit{temporal control}, which are discussed below. 

\textbf{Gesture Control: }Gesture control defines the fundamental static or dynamic gesture required to initiate a specific interaction, independent of spatial or temporal constraints. For example, to twist an object clockwise, the hand must first assume a grab gesture, then perform a rotation to complete the action. These gesture units build the foundation of a hand interaction. In this work, we demonstrate a core set of base gestures commonly used in virtual environment interactions, such as grab, pinch, and cut. For further technical details on extending the gesture set, please refer to \autoref{sec_sys}.

\textbf{Target Control: }A gesture can be either object-independent or object-dependent. For example, a user may perform a mid-air pinch gesture without interacting with any object (object-independent), or they may execute the same gesture on the corner of a virtual interface to manipulate it (object-dependent). The goal of target control is to determine whether a gesture depends on an object, identify the specific target the user is referring to, and adjust the hand movement accordingly to ensure proper interaction. For example, given the command ``grab the rightmost watermelon'', the target control should recognize this as an object-dependent action, resolve ``the rightmost watermelon'' as a specific object, and adjust the virtual hand's movement path to ensure it reaches and successfully grabs the correct target.

\textbf{Spatial Control: }Spatial control determines the movement and rotation of the hand in addition to the previous controls. This includes the movement of the hand in the 3D virtual environment and the rotation of the hand, including pan left/right, roll left/right, and tilt forward/backward. Beyond basic gesture and target control, spatial control modifies the movement path and behavior of the hand to enable more complex interactions. This allows users to perform tasks such as placing an object at a specific location, grabbing and moving a virtual slider, twisting a knob, or rotating the hand to inspect an item from different angles. 

\textbf{Temporal Control: }Temporal control acts as a playback control, regulating the timing and execution of hand interactions. This includes the ability to pause, resume, accelerate, decelerate an ongoing interaction, as well as undo, redo, or repeat a previous interaction. These controls enable interactions that depend on timing and states. For example, a user may pull the lever and stop pulling when they heard a audio cue to stop, or resize the window by pinching and moving to the right until satisfied. Users can also correct an undesired interaction through undo, or simplify repetitive interactions through repeating a specific hand state (e.g., do it again / 10 times). 

Given these control primitives, a hand interaction can be decomposed into one or more combinations of these foundational elements. As shown in \autoref{fig:primitives}, ``pull up'' can be broken down into a grab gesture (Gesture Control) and a upward movement (Spatial Control). {\m High-level commands that require multiple steps can be decomposed into multiple combinations of these primitives. For instance, ``put the apple into the basket'' involves three steps -- grab the apple, move to the basket, and release it.} Through this structured decomposition, the system can effectively interpret and reproduce a wide array of hand interactions, accommodating both simple and more complex interactions. This framework guides our system design as detailed in \autoref{sec_sys}, ensuring that our approach remains flexible and intuitive to various interaction needs.
\section{HandProxy: A Speech System for Virtual Environment Interaction Using a Proxy Hand}
\label{sec_sys}
HandProxy is a system that converts a user's live speech commands into dynamic virtual hand movements, allowing the proxy hand to interact with the virtual environment on the user's behalf. Overall, the HandProxy system has the following key capabilities:

\begin{enumerate}
    \item \textbf{Flexible natural speech commands: }HandProxy allows users to describe interactions in their own words without rigid formats. It interprets input, decomposes complex commands into executable steps, and prompts for clarification if needed. Users can issue commands at different levels of detail—from precise hand movements (e.g., ``pinch'' or ``grab the box'') to high-level goals (e.g., ``increase the brightness'' when a brightness knob is present), which brings additional flexibility for user input.
    \item \textbf{Context-aware command processing: }By leveraging environment metadata and real-time object positions, HandProxy interprets contextual references, allowing users to specify targets based on relative positions (e.g., ``grab the watermelon in the middle'') or object attributes (e.g., ``pick up the pink fruit''). It also maintains a history of interactions, enabling users to reference past actions (e.g., ``do it again'') or use undo/redo functions to restore a hand state, to support intuitive and efficient interaction.
    \item \textbf{Real-time streaming input and execution:} HandProxy takes speech input continuously, allowing users to speak naturally while the system interprets and executes commands in parallel for a seamless interaction experience. High priority commands (e.g., ``stop'', ``undo'') are processed instantly via rule-based methods, while LLMs handle complex instructions. System components run in parallel to minimize delays.
    \item \textbf{Feedback for disambiguation and system transparency:} To enhance clarity and transparency, HandProxy provides visual feedback overlays showing recognized commands, expected hand movements, and prompts for retry when needed. When multiple objects match the description in the command, it shows disambiguation labels to clarify user intent.
\end{enumerate}

\begin{figure*}[]
  \includegraphics[width=\linewidth]{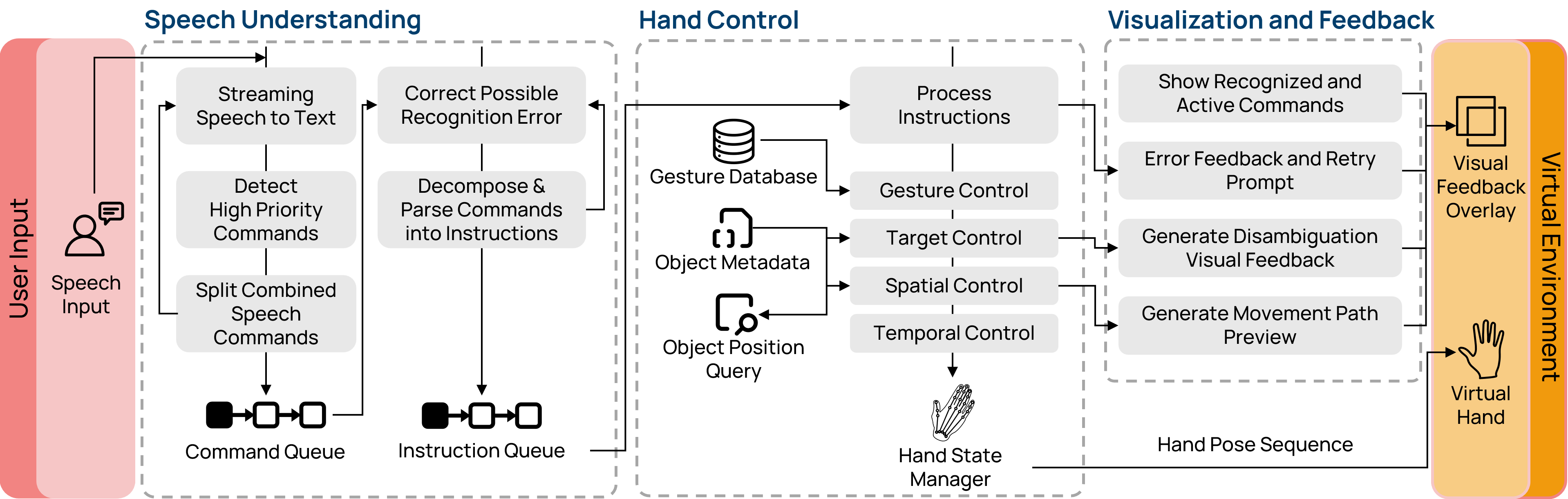}
  \vspace{-1.5pc}
  \caption{An overview of the HandProxy system structure. Given the user input, the speech understanding module recognizes the speech into commands, and parses the commands into system-interpretable instructions. The instructions are then sent to the hand control module, which generates the corresponding sequence of hand pose to control the movement of the virtual hand. Visualization and feedback module generates appropriate visual feedback for disambiguation and better transparency.}
  \Description{A flowchart illustrating HandProxy system consist of Speech Understanding, Hand Control, and Visualization and Feedback. On the left is the user input, which firstly goes through speech understanding module, processed as instructions, and then sent to the hand control. Hand control uses a gesture database to generate hand data, which is used to control the virtual hand in the virtual environment. }
  \label{fig:sys_structure}
\end{figure*}

{\m 
Below, we begin with an overview of the system design elements, followed by technical details of each components, and conclude with a summary of the design iterations conducted during development to offer rationale for our design choices.

\subsection{Design Elements}
As illustrated in \autoref{fig:sys_structure}, HandProxy consists of three major components: speech understanding, hand control, and visualization \& feedback. The speech understanding module continuously listens to and processes user input in a streaming fashion in real-time using Google Speech-to-text API \cite{google_stt}. It segments long speech inputs into individual commands based on punctuation, and adds them to the command queue. Using GPT-4o \cite{gpt4o}, the system then interprets each command and decomposes it into a sequence of instructions in the json format. Once instructions are available, the hand control module uses the hand pose sequences that was either retrieved from FPHAB \cite{fphab} dataset or recorded using LeapMotion and modifies the skeleton data to accommodate movement, rotation, and hand-object interaction. The playback of the hand data is controlled by the hand state manager to support interaction timing control, and is streamed to the target virtual environment for virtual hand control, along with additional visualizations and feedback to be overlaid on top of the user's view. The system components run in parallel to increase the efficiency and avoid blocking the hand movement and impacting user experience. 

The HandProxy system is implemented in Python, with its output streamed to a Unity application via the TCP connection. The Unity application renders a sample 3D virtual environment that contains interactable objects and a virtual hand, simulating an immersive application that supports hand gesture input. It overlays visual feedback specific to HandProxy on top of the immersive environment, as shown in \autoref{fig:sys_feedback}. For demonstration and prototyping purposes, the Unity app runs on a laptop, but can be deployed to other supported platforms (e.g., Android on Meta Quest). Demonstrations of the interaction experience on desktop and XR headset setups are provided in the supplementary video.
}

\subsection{Speech Understanding}
\label{sec-sys_speech}
The speech understanding module recognizes and interprets the user's command. It determines whether the speech command is an interaction command or a random speech, checks if it is ambiguous or interpretable, and decomposes it into structured, system-interpretable instructions for the hand control module to execute.

For speech recognition, HandProxy uses the Google Speech-to-Text API, which is configured to return intermediate results for real-time streaming recognition. The speech input is segmented into individual sentences, determined by terminal punctuation on the final results returned from the API. Each sentence is then treated as a command and is added to the command queue for interpretation. 

For command interpretation, HandProxy uses a multi-level approach for command processing. Similar to the reactive and deliberative layers in robotics \cite{airobotics}, the system uses keyword matching and directly responds to high-priority commands, with LLM-based processing for more complex inputs. Specifically, as soon as intermediate results are received from the speech recognition service, HandProxy checks predefined keywords and synonyms related to temporal controls as discussed in \autoref{par:building_blocks}. These commands --- such as ``stop,'' ``hold,'' or ``undo'' --- are time-sensitive and need to be executed with minimal delay. If a matching is detected, the system triggers the corresponding temporal control immediately. Otherwise, the system retrieves the earliest available command from the command queue and processes it using GPT-4o. The model is guided by a structured system prompt and is provided with a list of supported control categories from \autoref{par:building_blocks}, {\m which is listed below:

\begin{quote}{\footnotesize\ttfamily
Your objective is to break down a user command into actionable control components that the system can execute. Once you receive the command, follow these steps meticulously:

\#\#\#\# **Step 1: Command Fixing**

1. Input command is converted from speech, so minor errors may occur. If you notice any, please correct them and replace the original command with the fixed version.

\#\#\#\# **Step 2: Components Matching**

1. Carefully read through the entire command.

2. Imagine you are a person with only your right hand available (currently empty), standing in a wide space, surrounded by the objects and environment described in the command.

3. With point 2 in mind, align each command part (and necessary actions for execution, e.g., grabbing an object before moving it) with a corresponding control component listed below. Users may say or describe the interaction in various ways, the system should match the similar concept together. For example, "chop" and "cut" should both be matched to "cut", etc.

4. If no components can be matched, proceed DIRECTLY to Step 4.2, skip intermediate steps.

\#\#\#\# **Step 3: Ensuring Correct Order**

1. Components should follow the user's expected execution sequence.

\#\#\#\# **Step 4: Structuring Output**

1. Once the components are matched, fill in all fields within each component. ENSURE ALL FIELD VALUES ARE SELECTED FROM THE PREDEFINED OPTIONS PROVIDED BELOW.

2. Format each component as a JSON object as defined below, then output them sequentially in an array. If no components are matched, set the value of components to an empty array. 

[…\# a list of available controls and parameters for each component are appended after the prompt]
}\end{quote}}
The LLM is used to \textit{(i)} correct potential speech recognition errors (e.g., power bottom $\rightarrow$ power button), as used in other works like \cite{llm_asr_correct}, \textit{(ii)} decompose high-level commands, if any, into executable steps (e.g., put the cube into the basket $\rightarrow$ grab the cube, move to the basket, and release), and \textit{(iii)} parse individual steps as a list of instructions as json. Also, since people tend to use a surprisingly great variety of words to refer to the same thing \cite{vocab_problem}, the system prompt instructs the model to generalize beyond exact words provided in the supported instructions to accommodate diverse ways users may phrase their commands. For example, commands with similar intent are mapped to the same control action (e.g., ``pinch'' = ``tap index and thumb,'' ``pink circular fruit'' = ``peach''). If a given input is not relevant --- meaning it does not appear to be an interaction command --- the model is instructed to ignore it and return an empty list instead of attempting to interpret unrelated speech. Additionally, user commands are saved within a command list, allowing the system to refer back to previous inputs when necessary (e.g., resolving references to earlier commands). 

This approach ensures that the LLM interprets user commands and maps them to the appropriate hand interactions, including: 
\begin{itemize}
    \item \textbf{Gesture Control}: \texttt{{[``pinch'', ``point'', ``push'', ``grab'', ``swipe'', ``punch'', ``squeeze'', ``cut'', ``thumb\_up'', ``thumb\_down'', ``open\_hand'' (i.e., release)]}}
    \item \textbf{Target Control}: \textit{(i)} A list of interactable object names in the virtual environment, their current positions, and a list of descriptive tags for each object, \textit{(ii)} supported relative constraints to identify objects: \texttt{{[``below'', ``above'', ``to the left of'', ``to the right of'', ``in front of'', ``behind'', ``closest'', ``farthest'', ``first'', ``last'', ``on the left'', ``in the middle'', ``on the right'']}}
    \item \textbf{Spatial Control}: \textit{(i)} translational control: \texttt{{[``up'', ``down'', ``left'', ``right'', ``forward'', ``backward'']}}, \textit{(ii)} positional control: \texttt{{[``on top of'', ``under'', ``in front of'', ``behind'', ``to the left of'', ``to the right of'']}}, \textit{(iii)} rotational control: \texttt{{[``pan left'', ``pan right'', ``roll left'', ``roll right'', ``tilt up'', ``tilt down'']}}
    \item \textbf{Temporal Control}: \texttt{{[``stop'', ``continue'', ``faster'', ``slower'', ``undo\_step'', ``redo\_step'', ``hold'']}}
\end{itemize}

{\m At the end, the speech understanding module outputs a json object following this format. For example, }given the command ``pinch the cube,'' the system generates the following instruction:
\begin{quote}
    {{[\{``component\_type'': ``gesture'', ``value'': \{``gesture\_type'': ``pinch'', ``object'': ``cube'', ``is\_ambiguous'': False\}\}]}}
\end{quote}
Here, the system identifies a gesture control action, which specifies a pinch gesture on the cube object. If only one cube exists in the environment, the \texttt{is\_ambiguous} variable is set to \texttt{False}. However, if multiple cubes are detected, it is set to \texttt{True}, triggering a disambiguation mechanism and corresponding visual feedback in other modules to prompt the user for clarification.

For more complex commands, the LLM references both the list of supported actions and its general knowledge to determine the best way to combine supported hand controls to finish the user's request. For instance, given the command ``peach into the basket,'' the system generates multiple sequential steps:
\begin{quote}
    {{[\{``component\_type'': ``gesture'', ``value'': \{``gesture\_type'': ``grab'', ``object'': ``peach'', ``is\_ambiguous'': False\}\}, \{``component\_type'': ``movement'', ``value'': \{``movement\_type'': ``translational'', ``object'': ``basket'', ``is\_ambiguous'': False, ``position'': ``on top of''\}\}, \{``component\_type'': ``gesture'', ``value'': ``release''\}]}}
\end{quote}
By dynamically generating one or more executable steps, HandProxy lets users issue both low-level direct commands and high-level intent-based commands, which provides an adaptive and natural interaction experience.

\subsection{Hand Control}
The hand control module executes instructions from the json to generate a sequence of hand pose data to control the virtual hand. It maintains a hand state manager, which tracks key hand properties such as the current hand pose, wrist position, playback status (e.g., whether an action is in progress), and the sequence of upcoming hand poses. When a new instruction arrives, the hand control module activates the relevant control sub-modules discussed in \autoref{par:building_blocks} to update the hand pose data. Next, we describe each control module in detail.

\subsubsection{Gesture Control}
The gesture control module generates a sequence of hand pose data to represent the required gesture, which are retrieved from the FPHAB \cite{fphab} dataset or captured using LeapMotion, but is extensible to gesture datasets that include the 2.5D coordinates \texttt{(x\_ratio, y\_ratio, z\_depth)} of 21 hand joints. Currently supported gestures include grab, pinch, point, push, swipe, cut, punch, squeeze, thumb up/down, and open hand (release).

To enrich data with additional gesture information, each gesture is manually annotated with details about the gesture’s object dependency and different gesture stages \cite{hand_and_mind} that could benefit the gesture control. {\m The full metadata format can be found in \autoref{ap:gesture_metadata}, and }some key attributes include:
\begin{enumerate}
    \item \textbf{Is static: }Specifies whether the gesture is static or dynamic.
    \item \textbf{Interacting frame: } Identifies the frame at which the hand fully interacts with an object (if applicable).
    \item \textbf{Interacting joint: }Specifies the joint primarily involved in the interaction (e.g., the index fingertip for a pointing gesture). If unspecified, the wrist is used by default. This information helps hand movement calculations during object interactions to better match the expected hand behavior.
    \item \textbf{Segments: }Defines the start and end frames for different gesture phases, including preparation, stroke, and retraction, to support additional features such as gesture holding.
\end{enumerate}

{\m Combining all the data together, when given a gesture command, the module loads the sequence of hand pose data from the gesture database. Downstream modules can use the pose data at each frame as an $21\times3$ matrix, modify it, and send it to the virtual environment.}

\subsubsection{Target Control}
The target control module handles object-dependent hand interactions by identifying and aligning the virtual hand with the intended target. It first resolves the target object using positional constraints, such as ``below,'' ``above,'' ``to the left of,'' ``to the right of,'' ``in front of,'' ``behind,'' ``closest,'' ``farthest,'' ``first,'' ``last,'' ``on the left,'' ``in the middle,'' and ``on the right.'' These constraints are resolved by retrieving the current positions of all virtual objects through a query to Unity application and calculating their relative spatial relationships. 

Once the target object has been determined, the module calculates a path to move the hand from its current position to the target object. {\m It first determines the total distance to move, calculates the movement step size by dividing the total distance by the number of frames for this gesture, and cumulatively add step sizes to original hand pose data for all the following gesture frames. }This calculation also accounts for the interacting joint (as discussed in Gesture Control) to ensure proper alignment of the hand with the object. For example, if the command is to point at a button, the system uses the interacting joint defined in the gesture metadata to set the index fingertip to align with the center of the button when calculating the movement path. 
By dynamically modifying the original hand pose sequence, the target control module ensures that gestures interact with objects in the correct fashion.

\subsubsection{Spatial Control}
The spatial control module manages the movement and rotation of the virtual hand. It supports translational movements in six directions: ``up,'' ``down,'' ``left,'' ``right,'' ``forward,'' and ``backward.'' Additionally, it enables rotational adjustments, including ``pan left,'' ``pan right,'' ``roll left,'' ``roll right,'' ``tilt up,'' and ``tilt down.'' Specifically, translational movements are applied by adding a speed vector to the hand pose data, while rotational adjustments are calculated by rotating the joint coordinates along the estimated hand axis using the vector of middle metacarpophalangeal joint (MCP) to index MCP, middle MCP to wrist, and the vector perpendicular to these two vectors.

Similar to target control, movement commands can also reference target objects along with relative position constraints such as ``on top of,'' ``under,'' ``in front of,'' ``behind,'' ``to the left of,'' and ``to the right of.'' In such cases, the system computes the movement path similarly to object interactions but applies an offset to reach the specified position relative to the target object.

\subsubsection{Temporal Control}
Temporal control specifies the timing and state changes of hand interaction, including ``stop'', ``continue'', ``faster'', ``slower'', ``undo\_step'', ``redo\_step'', and ``hold.'' Hand pose data from earlier stages is sent to the hand state manager, which keeps a playback speed (frames per second, fps) and playback state (stop, play) that can be adjusted based on temporal control commands. For the hold command, specifically for holding a specific gesture (e.g., as in grab and hold the cube), the temporal control adds a key frame at the interacting frame (as defined in gesture control) and pauses the playback at that frame to hold the gesture.

Additionally, the hand state manager maintains a history of hand poses after each command. This enables ``undo'' and ``redo'' functionality, allowing users to revert to previous hand states or redo an action if needed. If an error occurs, the system can quickly restore the last hand state, providing a flexible interaction experience.

{\m At the end, the hand control module outputs the hand pose for the current frame as a list of 63 numbers (21 joints $\times$ (x, y, z)), and sends it to the Unity app to control the virtual hand.}

\subsection{Visualization and Feedback}
\begin{figure*}[]
\includegraphics[width=.98\linewidth]{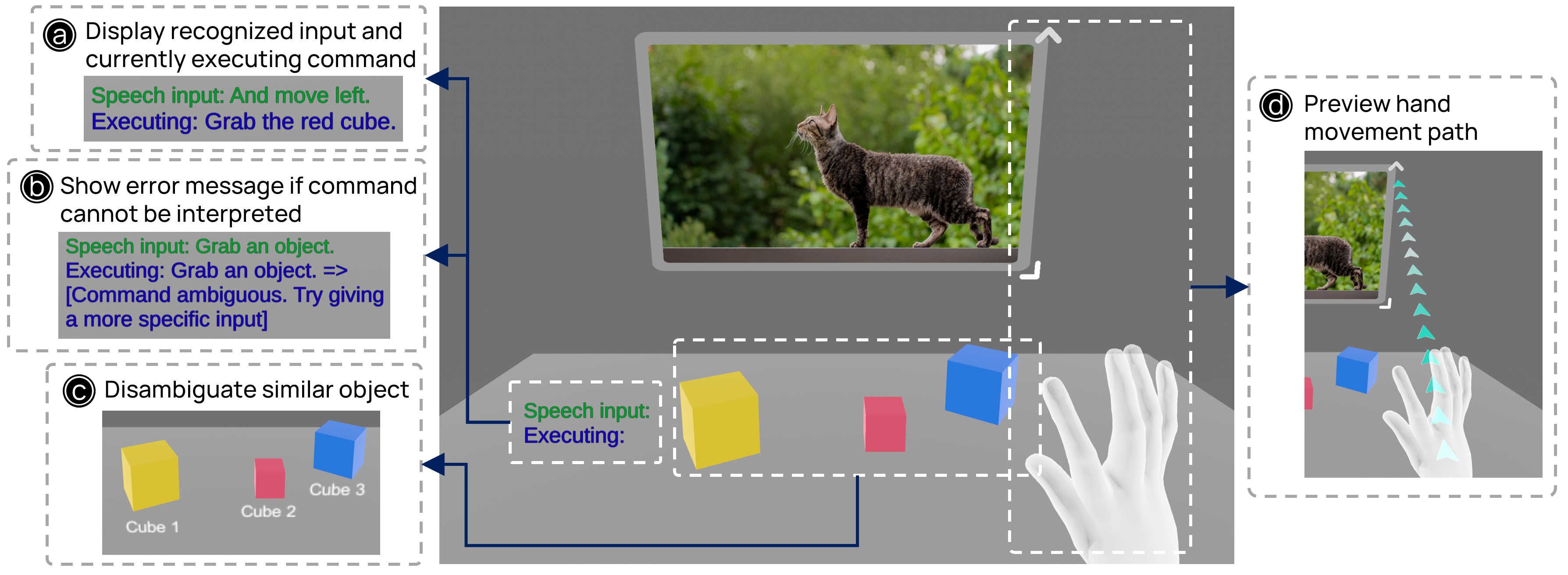}
  \vspace{-.5pc}
  \caption{{\m Examples of system feedback: (a) an overlay streaming speech recognition results and currently executing command, (b) an error message indicating the command is ambiguous, prompting the user to try again, (c) a disambiguation overlay that labels ambiguous objects, where the user can clarify the desired one by saying its number, and (d) a preview of the hand movement path, shown as a list of arrows pointing towards the destination.}}
  \Description{A figure showing 4 different visual feedback, where the first and second one overlays the speech results, executing commands, and error messages as text on the lower left corner of the figure. Disambiguation feedback adds a small text label with numbers under each duplicate objects (e.g., multiple cubes). The movement preview shows a line of arrows pointing to the movement direction. }
  \label{fig:sys_feedback}
  \vspace{-1pc}
\end{figure*}

To enhance system transparency and reduce ambiguity, HandProxy uses various visual feedback mechanisms that are overlaid on the user’s viewport. {\m In the prototype Unity app, visualization is implemented as a canvas overlaid on the main camera view. Each visualization is triggered when it receives the command from the HandProxy. \autoref{fig:sys_feedback} shows examples of each feedback type.} To ensure users understand how their commands are processed, the system displays both the recognized (full input) and active (which part of the input is being executed) commands. If a command is irrelevant (e.g., not a valid hand interaction) or cannot be interpreted, the system highlights an error message in the recognized input and prompts the user to rephrase the command. For cases where the user refers to an ambiguous object (i.e., multiple similar objects exist in the environment), the system provides disambiguation hints by overlaying numbered labels on each possible target, allowing users to specify the object by its assigned number. To further increase transparency in hand movement execution, the system visualizes the movement path. As shown in \autoref{fig:sys_feedback}d, a sequence of arrows appears along the trajectory, indicating the direction in which the hand is moving. If the command involves multiple steps, the full movement sequence is previewed to give users an overview of how the hand will execute the task. These visualizations provide users with greater visibility of the system to bridge the gulf of evaluation, improving both command accuracy and interaction efficiency while reducing errors and misinterpretations.

{\m \subsection{System Latency}
The system latency is composed of two main components: speech recognition and processing latency, and command processing latency. Speech recognition and processing latency measures the time from just before an audio chunk is sent via an API request, to when the recognized text is split into commands. Based on 100 requests, the average latency for this process was 0.18 seconds, with a standard deviation of 0.07s. The command processing time is the time it takes for the system to interpret the command and generate a sequence of corresponding hand pose keypoints. We measured such latency for both commands that matched high-priority keywords, and those processed by LLMs. Through a test of 100 commands that matched the high priority keywords, all commands were executed within 0.001 seconds. Through a test of 100 requests that included both simple (e.g., requires one step), complex (e.g., needs to be decomposed into multiple steps), and irrelevant commands (e.g., not a valid interaction instruction), the average response time was 1.39 seconds, with an standard deviation of 0.81s.}

{\m \subsection{Adapting to New Gestures and Environments}
HandProxy can be extended with additional gestures. It accepts hand pose data in the format of a sequence of 21 keypoint coordinates (x, y, z) and a gesture metadata in the format shown in \autoref{ap:gesture_metadata}. The widely used 21-joint format enables data from various sources to be easily integrated, such as hand pose datasets \cite{fphab}, recorded data (e.g., from LeapMotion or MediaPipe), and hand pose generation models \cite{text2hoi}, and HandProxy automatically normalizes the data. Existing built-in controls, such as target, spatial, and temporal controls and other gestures, are still compatible with the newly added gestures. 

To adapt to new environments, HandProxy requires the name and optionally descriptive tags of interactable objects, similar to the accessibility metadata for UI elements on mobile \cite{android_metadata} or web interfaces (e.g., DOM tree \cite{dom}). The GPT prompt is dynamically updated with the object information to process new commands. Note that due to limited direct access of the hand input on XR systems, in our current implementation, the virtual hand is added as an object within the application. However, we envision that with system-level support, HandProxy can directly control the virtual hand input of the operating system and retrieve interactable object information to achieve a more unified, cross-application control.
}

\subsection{Design Iterations}
The HandProxy system was iteratively designed and developed through pilot studies with 13 participants {\m recruited from the student group at the authors' institution. Participants were asked to use HandProxy to perform object selection, manipulation, and transformation tasks in a test virtual environment, and is then interviewed for their experiences and potential improvement suggestions.} Here, we summarize key aspects from these iterations, with the goal of providing additional rationale behind our design choices and insights into user preferences. 

\subsubsection{Enhancing the Speech Interface}
The speech interface is a core component of HandProxy. In our initial design, we followed a traditional turn-based interaction approach, similar to off-the-shelf speech interfaces \cite{apple_voice_control}, where users issue a command, wait for execution, and then issue the next command. However, our pilot studies revealed that this approach was inefficient for hand control. Many participants preferred to issue multiple commands continuously, describing an entire sequence of actions in one go. The forced pauses between commands disrupted the interaction flow, making it less natural. Additionally, it was difficult for users to issue commands to override current executions, such as stop or undo, as users had to wait for the previous command to finish before issuing a correction. As a result, we redesigned the speech interface to support continuous input, allowing simultaneous speech recognition and action execution. This change also motivated the parallelized system architecture, where each module runs in a separate thread, ensuring that speech processing, command interpretation, and execution run without blocking other processes.

Additionally, we observed variations in user preferences for the level of detail in commands. Some preferred step-by-step instructions, such as ``move left,'' ``grab,'' while others issued higher-level commands, such as ``put the cube into the basket,'' expecting the system to infer intermediate steps. {\m This motivates the choice of integrating LLMs (GPT-4o) than pre-defined commands to support flexible and diverse speech input and leverage their reasoning capabilities to interpret user intent dynamically.}

\subsubsection{Integrating Contextual Information}
Throughout our iterations, we observed that participants naturally used descriptive references to identify target objects, including shape, appearance, and relative location. 
Examples include ``green ball'' or ``the object next to the button.'' This demonstrated the need for contextual understanding, which motivated our decision to incorporate object metadata into the system.

While advanced vision models could enhance contextual understanding, we drew inspiration from accessibility metadata in 2D user interfaces and implemented a metadata structure for virtual objects. Each object is assigned a set of descriptive tags, allowing the system to recognize and differentiate objects based on user descriptions. Additionally, many object references rely on common knowledge (e.g., ``a green fruit with stripes'' is likely a watermelon). To enhance generalizability, we integrated LLMs to allow HandProxy to reason beyond predefined metadata and infer likely object references based on common knowledge and context.

\subsubsection{Improving System Transparency}
Speech recognition errors are a common challenge, and users wanted to know how their commands were interpreted. This feedback motivated several transparency focused design choices, including showing the recognized results and visualizing the path of the expected hand movement. These features ensure that users can intervene in real-time when error happens using commands like stop or redo, which ultimately improves interaction reliability and usability.

\section{User Study}
We conducted a user study with 20 participants to evaluate the HandProxy system and gather insights on usage patterns and feedback. Specifically, our research questions are as follows:

\noindent\textbf{RQ1. } To what extent can participants complete a wide variety of tasks using HandProxy?

\noindent\textbf{RQ2. } How effectively can the system interpret the diverse ways in which users issue commands?

\noindent\textbf{RQ3. } What strategies, preferences, and methods do participants use for specifying intent?

\noindent\textbf{RQ4. } How do participants experience and perceive HandProxy?

\noindent\textbf{RQ5. } What other expectations do participants have for the system?

\subsection{Participants}
We recruited 20 participants (9 female, 11 male, age 18-31) from the student population at our institution. Based on a self-reported survey, 3 participants had no prior experience in immersive environments (e.g., VR/AR), 13 were beginners, and 4 were intermediate users. For experience with hand-gesture-controlled interfaces (e.g., XR headset like Meta Quest or hand tracking cameras like Kinect), 5 participants reported no experience, 12 were beginners, and 3 were intermediate users. Regarding speech interfaces, 1 participant had no experience, 5 were beginners, 12 were intermediate users, and 2 were experts. Examples of speech systems included virtual assistants like Siri or Alexa, ChatGPT voice mode, and speech input on TV remotes. The study was approved by the IRB, and participants received \$30 gift cards as compensation.

\begin{figure*}[]
  \includegraphics[width=.98\linewidth]{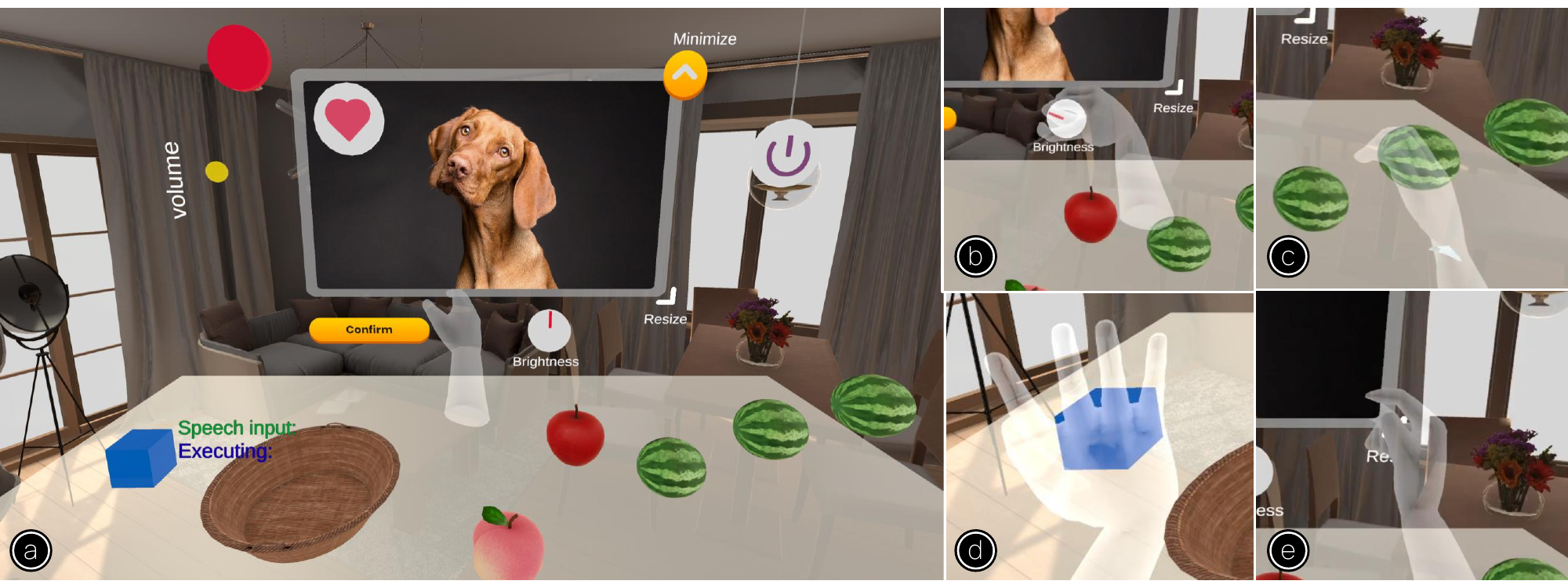}
  \vspace{-1pc}
  \caption{Virtual environment for the user study (a), with different interactable 3D objects and UI widgets. Participants were asked to perform various hand interactions by commanding the virtual hand in the environment. Task examples include but not limited to twist knob (b), grab fruit (c), push cube (d), and pinch to resize window (e).}
  \Description{A figure of the virtual environment participants used in study. The environment has a 2D image viewer in the center, a table below it with apple, peach, three watermelons, cube, and basket. Near the viewer there is a volume slider on the left, confirm button, brightness knob, resize button below it, minimize button above it, and a power button to its right. The virtual hand can be used to perform different tasks on them, such as twist the brightness knob, grab watermelon, push cube, and pinch resize button.}
  \label{fig:us_env}
\end{figure*}

\subsection{Apparatus and Procedure}
The study sessions were held in a small meeting room. Participants sat in front of a TV connected to a workstation laptop with i9-12900H CPU, {\m which runs the 3D virtual environment in Unity. }This setup is to provide an equivalent experience for all participants. As shown in \autoref{fig:us_env}, the virtual environment in Unity is designed to have different 3D objects and UI widgets, supporting a variety of interaction tasks. {\m The interactions are triggered only by collision or gesture detection of the virtual hand. The Unity app does not provide API to trigger interaction.} Each study session took 2 hours. 

\textbf{Overview (10 minutes):} participants were broadly introduced to the project and the system setup, and were asked to fill out the pre-survey on their prior experiences.

\textbf{Study tasks (60 minutes)}: for each task, participants first watched a video clip demonstrating a virtual hand performing the intended hand interaction. They were then instructed to replicate the interaction using speech through HandProxy. To minimize bias toward specific commands, no guidance was provided beyond the video demonstration, except when participants required clarification or assistance. For mid-air gesture tasks, participants were asked to complete the task directly. For other tasks, interactions were conducted with three different objects. During the initial \textit{practice session}, participants used the first object to explore the system’s capabilities and limitations by experimenting with different commands. In the subsequent \textit{test session}, they performed the same interaction on two additional objects, aiming to complete the task as accurately as possible. The tasks below are selected based on common types of interaction tasks in the virtual environment \cite{hands_free_VR}, including selection, manipulation, and transformation.
\begin{enumerate}
    \item \textit{Warm-up with mid-air gestures: }participants were asked to use speech to reproduce common mid-air gestures on commercial XR devices \cite{apple_gestures, quest_gestures} as a warm up, which included pinch, swipe left, double pinch, and thumb up. They served to familiarize participants with the system and the basic gestures, and to prepare and on-board them for subsequent tasks.
    \item \textit{Hand interactions: }participants were asked to perform a wide range of interaction tasks for virtual environment \cite{hands_free_VR}, including object selection, manipulation, transformation. The tasks included object dependent gestures (grab \{apple, peach, blue cube\}, press \{confirm, minimize, power\} button, pinch \{blue cube, volume slider, resize button\}, push \{confirm, like, power\}) and disambiguation (grab the \{left, middle, right\} watermelon among the 3 watermelons). 
    \item \textit{Movement and rotation control: }this included hand movement (move \{left, up, forward\}), object movement (put the \{apple, peach, blue cube\} into the basket, rotation (turn the brightness knob \{clockwise, counterclockwise\}). Participants were also asked to perform tasks that required timing control. This included grab the \{apple, peach, cube\} and move to the left according to a light signal (green = move, red = stop), and to press and hold the \{power, like, confirm\} button. 
    \item \textit{Complex tasks: }for these tasks participants were only shown the image of the final outcome, without demonstrating the steps to accomplish it. It was up to the participants to decide what and how many steps to take to finish the task. These include tasks to make the window wider, put apple, peach and first watermelon into the basket, and maximize the volume.
\end{enumerate}

\textbf{Free exploration (10 minutes): }participants had the opportunity to freely explore the environment using commands of their choice to investigate the system's capabilities and limitations. A Likert scale questionnaire on system usability was given at the end. 

\textbf{Semi-structured interview (40 minutes): }participants were interviewed about system performance, user experience, interaction strategies, desired features, and suggested improvements.

{\m
\subsection{Methodology and Procedure of Data Collection and Analysis}
To support quantitative analysis, the system logged the recognized text, decomposed json output, executed hand controls, and system feedback (e.g., visualization, error messages) for each of the command user gave during the study. For commands that were incorrectly recognized, the researcher noted down the original command for analysis. For commands that were incorrectly interpreted or executed, the researcher marked the specific task and used the system log for further analysis. The command inference time was also recorded to evaluate the system responsiveness. To focus on task-related input, we excluded user input that were not relevant to the task they were asked to perform. This could be due to misunderstanding of the task goal, or accidental misspoke commands. At the end, a questionnaire with Likert scale questions were used to evaluate the system usability. 

For qualitative analysis, we audio-recorded semi-structured interviews with participants' permission, and conducted a thematic analysis \cite{thematic} on the transcripts. The primary researcher generated the initial codes and themes, and collectively discussed and examined the results with the research team to reach consensus. 
}

\section{User Study Results}
Below we present the study results. Unless otherwise noted, the findings are based on data from the test sessions.
\subsection{RQ1: To what extent can participants complete a wide variety of tasks using HandProxy?} 
\label{rq1}
\textbf{Participants successfully completed various tasks using HandProxy,} with an 100\% task completion rate on tasks conducted in the test sessions. Among the 781 commands participants issued across all test sessions, the system correctly executed 717 of them, with an overall accuracy of 91.8\%. The median command execution accuracy per participant is 92.5\%, with an IQR of 89.3\% to 95.5\%. {\m On average, the system took 1.66 seconds to interpret the recognized speech command, with an standard deviation of 0.94 seconds. }

After errors happened, \textbf{participants were able to recover from errors by using alternative commands to complete the task}. Multiple techniques were reported for error recovery, including rephrasing (13), repeating (3), and splitting into smaller steps (3). Across all commands in the test sessions, an average of 1.09 attempts {\m (std 0.33)} were needed for a command to succeed. Specifically, 85.5\% of error commands were resolved with 1 more attempt, 12.7\% took 2 more attempts, and 1.8\% took 3 more attempts, showing that most errors were corrected with one alternative attempt.

\subsection{RQ2: How effectively can the system interpret the diverse ways in which users issue commands?}
\begin{figure*}[!hp]
    \centering
         \includegraphics[width=\linewidth]{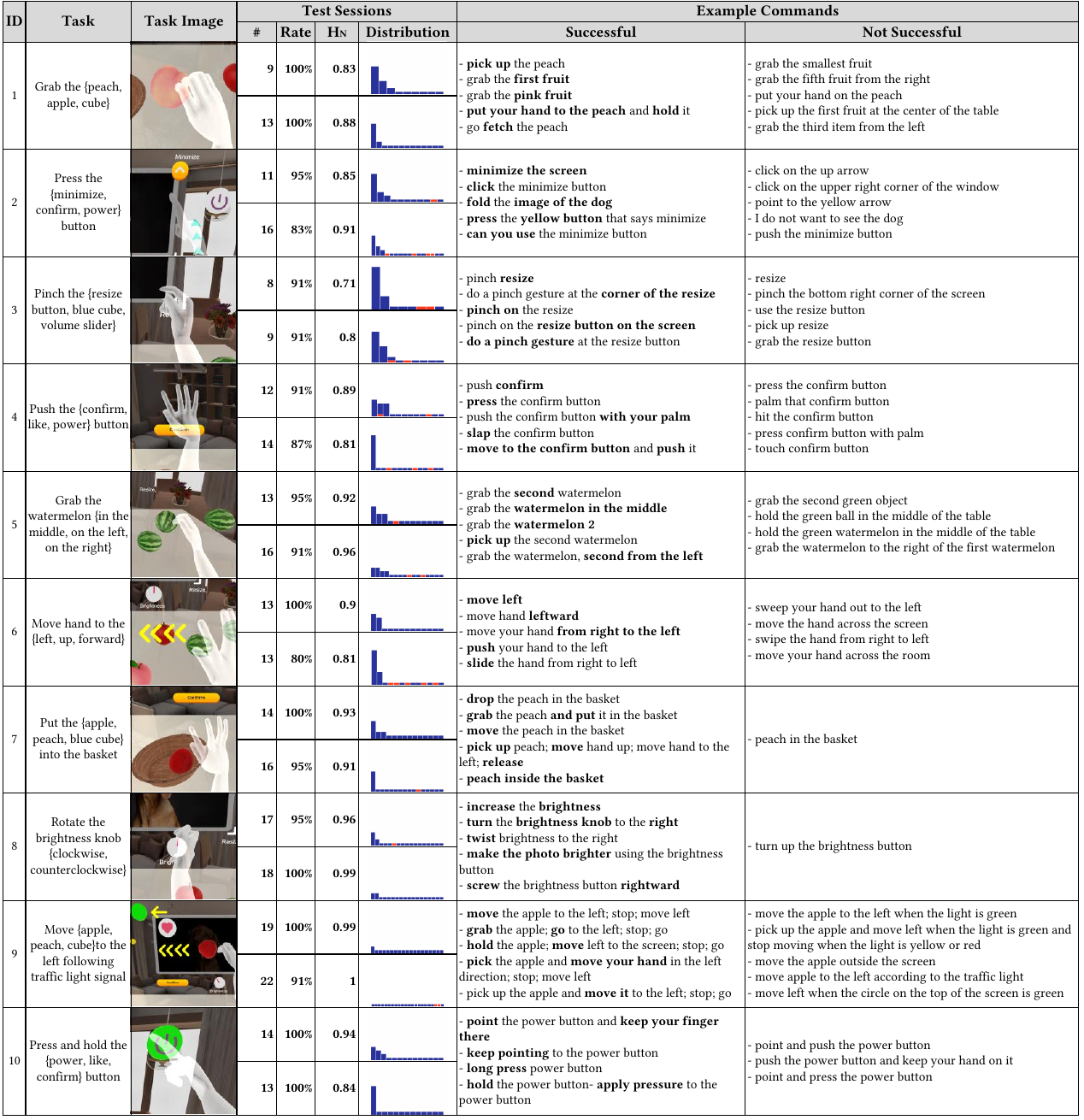}
    \caption{Command diversity in test sessions: Each task was repeated on three objects --- one for practice and two for test sessions (top row: second object, bottom row: third object). We report the number of unique command phrasings (\#), execution accuracy (Rate), normalized entropy ($H_{N}$), {\m and a histogram showing the distribution of unique phrasings, sorted by frequency. We color-coded the successful attempts as blue and unsuccessful attempts as red to illustrate the distribution of correct/incorrect commands.} Finally, we provide examples of successful and unsuccessful participant commands. }
    \Description{A table of the name, number of unique commands, success rate, normalized entropy, distribution, successful, and unsuccessful commands for each study task. }
    \label{fig:us_diversity}
\end{figure*}

\textbf{HandProxy effectively handled users' diverse ways of phrasing commands, including less common ones.} To identify unique commands per task, we lemmatized and removed stopwords from commands used by participants during test sessions. \autoref{fig:us_diversity} shows the histogram and normalized entropy of unique commands participants used to complete study tasks. The normalized entropy quantifies the variation in the distribution of commands with a value from 0 (all attempts were the command) to 1 (each attempt corresponds to a unique command). Both the histogram and normalized entropy shows the diversity of commands participants used. Specifically, tasks were not dominated by just one command but were distributed across various commands, and the system was able to interpret most commands, including those used less frequently.

During practice sessions, participants generated an average of 32 unique commands per task. The system successfully interpreted command variations, including different verbs (e.g., \{grab, pick up, fetch, hold\} the peach), object descriptions (e.g., peach, pink fruit), spatial references (e.g., first fruit, second watermelon from the left), and sentence structures (e.g., can you use the minimize button, peach inside the basket). Moreover, the system demonstrated the ability to interpret and decompose high-level commands and generalize beyond literal meanings. For example, it interpreted ``increase the brightness'' as \textit{(i)} grab the brightness knob and \textit{(ii)} twist right. In the example of ``fold the window,'' the system was able to connect fold to minimize, and press the minimize button.

After practice sessions, as shown in \autoref{fig:us_diversity}, participants continued to use diverse commands rather than relying solely on the most straightforward ones, and the system effectively handled this variability. Given that participants were free to use their own words without predefined commands or formats, these results highlight HandProxy's capability to interpret flexible and varied input.

To better understand HandProxy's limitation, we analyzed the retried commands in the test sessions. In total, \textbf{the system failed to execute or incorrectly executed 64 commands}. Among them, 40 {\m valid commands were categorized as invalid, and the system prompted users to try again and did not perform any action}. Most were due to challenges {\m for the LLM in} identifying target objects (n=26), including synonyms (9), visual descriptions (6), object functionality (6), positional constraints (3), or other system errors (2). For example, participants described the power button as ``the white and round button on the right side of the screen'' (P1), the basket as ``the brown object'' (P16), or directly described the visual content within a widget, such as describing the window as ``the dog photo'' (P2). In other cases, participants referred to objects with synonyms or their expected functionality, such as describing the power button as ``start on off button'' (P10), or the heart button as ``favorite button'' (P16). While in many cases the system could infer beyond the object metadata, it struggled to account for the possible descriptions described above. This shows the need for a more comprehensive understanding of the objects and environment to support more diverse descriptive commands, which we will discuss in \autoref{sec:discussion}. 

Apart from invalid commands with no action, the system incorrectly executed 24 valid commands. Among them, 3 were due to speech recognition errors, while 18 resulted from misinterpreted hand interactions. These include ambiguous commands (6), incorrect command decomposition (5), incorrect gesture parsing (4), or performed on the wrong target objects (3). Some commands had multiple possible interpretations, where context could have clarified the intended meaning. For instance, when P6 said, ``turn the volume slider to maximum,'' the participant expected the slider to move to the top. However, the {\m LLM} interpreted the word ``turn'' literally and {\m the system} did a ``twist right'' gesture instead. Errors also occurred when participants combined multiple commands into one sentence. For example, P20 said, ``pinch the resize button and pull it to the right'', expecting a continuous pinch-and-move gesture. Instead, the {\m LLM} parsed ``pull'' as a separate command, initiated a ``grab'' gesture that canceled the pinch and caused an unexpected action. The system identified the wrong object from some other commands. For example, P5 said ``pick up the first watermelon on the right'' to refer to the rightmost watermelon, but the system interpreted it as the first watermelon overall (left to right).

Nonetheless, participants were able to complete 100\% of the tasks and recover from errors using alternative commands, with an average of 1.09 attempts {\m (std 0.33)} per command. 

\subsection{RQ3: What strategies, preferences, and methods do participants use for specifying intent?}
\label{rq3}
\textbf{Participants demonstrated diverse strategies and preferences while prompting the system. }
In practice sessions, the average command length was 5.25 words {\m (std: 2.72, median: 5)}. In test sessions, the average was slightly lower at 4.73 words {\m (std: 2.32, median: 4)}. For free exploration sessions -- where users were not restricted to specific tasks -- the average increased to 5.95 words {\m (std: 3.32, median: 5)}. Participants who preferred more detailed commands found it easier to ``specify what I want'' (P2) or believed it would ``help the system understand'' their intent more accurately (P7). For example, to grab the peach, P15 used the command, ``can you grab the peach and hold it in your hand?'' In the interviews, 14 out of 20 participants favored shorter commands for their simplicity, clarity, and lower risk of errors. As P16 mentioned, ``The longer it is, the more words I could say wrong and it could misinterpret.'' 

However, having shorter commands does not necessarily mean commands are always simple and low-level. In fact, participants used a mix of high-level and direct commands, such as ``maximize the volume'' (high-level) or ``pinch the volume slider'' followed by ``pull up'' (direct). 
For example, P9 mentioned that ``I just wanted to tell the system what I want to do, and let the system figure out what gestures to do,'' while others mentioned that using detailed control would be ``more accurate'' (P8), especially in precise tasks (P11) or those that need fine-grained controls (P6). Participants also reported the need to switch to a new mental model while using the system. This led some participants to use more descriptive ways of specifying their interaction intentions, such as ``touch your index finger and thumb'' to describe the pinch gesture (P13). As P17 said,
\begin{quote}
    \textit{``A lot of the motions that you intuitively perform didn’t come to me in words very easily. So my strategy was to describe exactly what the hand was doing until I’d sometimes realize, `Oh, that’s how it should be [described]' ''}
\end{quote}

\subsection{RQ4: How do participants experience and perceive HandProxy?}

We evaluated system usability using 7-point Likert scale questions, covering system effectiveness, interaction variety, ease of use, responsiveness, learnability, consistency, and user confidence. {\m The questions are based on the System Usability Scale \cite{sus}, with modifications to better align with the specific tasks in this user study. The Likert scale questions were administered three times at different stages to identify potential learning effects. However, no significant differences were found between them. Therefore, we report the results from the final administration that were conducted after all tasks were completed, as they provide the most comprehensive reflection of participants’ overall experience with the system.}

Overall, participants found HandProxy effective for hand interactions (avg. 5.5, {\m std. 0.82}), can support various interaction gestures (average 5.75, {\m std. 0.91}), easy to use (average 5.7, {\m std. 1.12}), and participants were generally confident in using it (average 5.3, {\m std. 1.08}). Regarding system responsiveness, participants gave an average of 5.15 ({\m std. 1.18}). While participants appreciated the fast timing controls (e.g., stop, continue, undo) (e.g., P6), overall responsiveness across different types of commands could be improved for smoother interaction (P2, P8). HandProxy was also seen as intuitive and easy to learn (avg. 5.7, {\m std. 1.41}), especially given that the study was intentionally designed to have only a very brief on-boarding session. As P3 mentioned, ``a few minutes of play and you should be ready.'' However, P15 highlighted challenges for users unfamiliar with AR/VR, often relying on long, descriptive commands and struggling with terminology. Both P15 and P11 suggested that a tutorial with example commands, words, and objects would make the system more intuitive for beginners.

Additionally, participants valued HandProxy's ability to handle high-level commands (P1, P2, P8, P11, P19); robustness to certain ambiguity in the command (P5, P10, P11); consistency on the same task (P7, P15); its memory capability (P19); and its ability to take continuous speech input and perform actions on the go (P7). For example, P1 mentioned that HandProxy is able to directly execute her high-level commands step by step, ``even though I haven't told the system how to perform the actual task.'' 

\begin{figure}[]
    \centering
         \includegraphics[width=\linewidth]{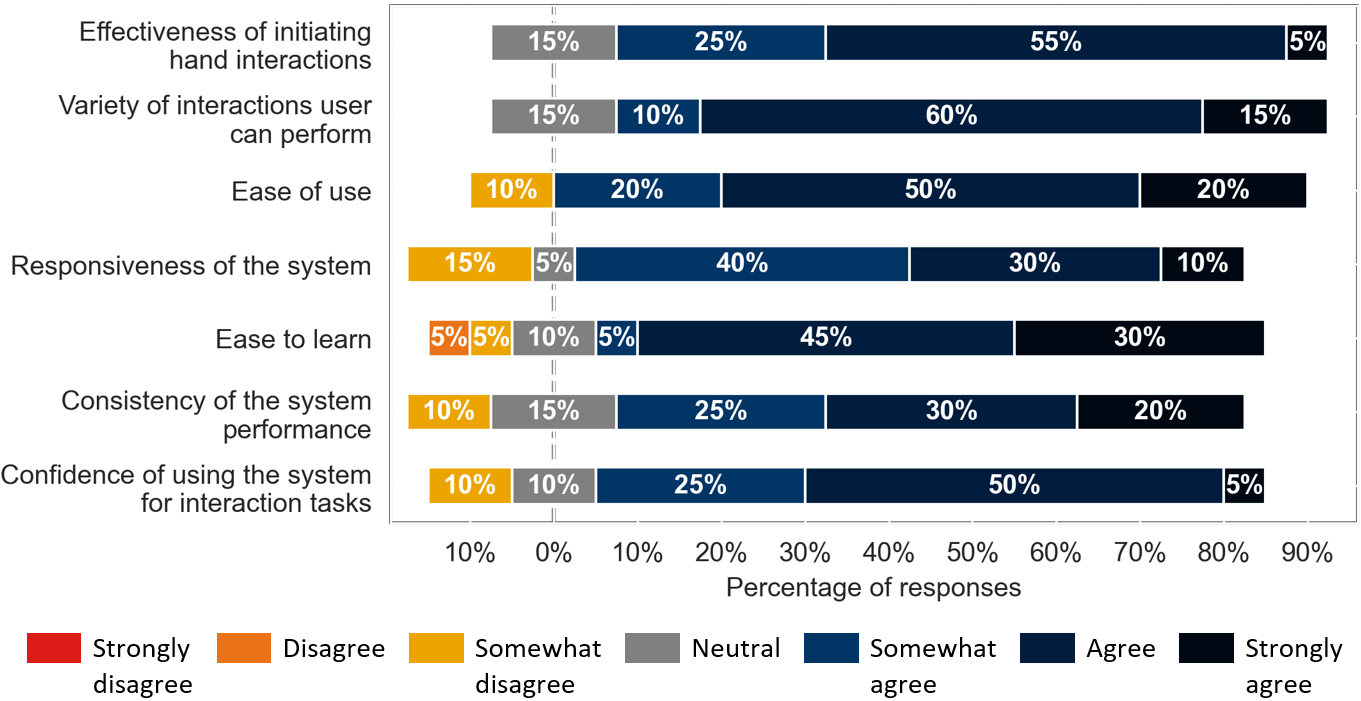}
         \vspace{-.5pc}
    \caption{{\m Final Likert scale responses of system usability.}}
    \Description{A figure of the final Likert scale responses across seven evaluation criteria related to the system's usability and performance. The categories on the vertical axis include: Effectiveness of Initiating Hand Interactions, Variety of Interactions User Can Perform, Easiness of Use, Responsiveness of the System, Easiness to Learn, Consistency of the System Performance, Confidence of Using the System for Interaction Tasks. The majority of responses lean towards agreement and strong agreement, particularly in categories of "Easiness of Use," "Easiness to learn," "System consistency," and "System variety". Specifically, for Effectiveness of initiating hand interactions, majority agreed (55\%), strongly agreeing (5\%) or somewhat agreed (25\%), with 15\% neutral. For Variety of interactions user can perform, 60\% agreed, 15\% strongly agreed, 10\% somewhat agreed, 15\% neutral. For Ease of use, 50\% agreed, 20\% strongly agreed, 20\% somewhat agreed, 10\% somewhat disagreed. For responsiveness of the system, 40\% somewhat agreed, 30\% agreed, 10\% strongly agreed, 15\% somewhat disagreed, 5\% neutral. For Ease to learn – 45\% agreed, 30\% strongly agreed, 10\% neutral, and 5\% each for somewhat agree, somewhat disagree, and disagree. For Consistency of the system performance, 30\% agreed, 25\% somewhat agreed, 20\% strongly agreed, 15\% neutral, 10\% somewhat disagreed. For Confidence of using the system for interaction tasks, 50\% agreed, 25\% somewhat agreed, 5\% strongly agreed, 10\% each for neutral and somewhat disagreed.}
    \label{fig:us_likert}
    \vspace{-1pc}
\end{figure}

Interestingly, we found that \textbf{participants had different perceptions of the virtual proxy hand -- either as an interaction tool, an agent, or a part of their body.} These not only impacted the way they used it, but also their expectations about the system's capabilities. Participants who perceived HandProxy as a tool treated the virtual hand more like a cursor, and cared more about what it could achieve than how realistic the interaction was, and in some cases, suggested that it should go beyond what a real hand can do. As P11 mentioned,
\begin{quote}
    \textit{``I don't know if the system can make the hand more abstract. For example, it could hold many things at a time. It may not need to be limited to the physical world, allowing for actions beyond what we can do in real life.''}
\end{quote}
This perception of HandProxy as a tool also influenced how participants talked to the system. For example, P7 used simple, direct, ``non-human speech'' to complete tasks,  ``just like other voice controls that's relatively old.'' This perception may have led some participants to underestimate the system’s ability to handle high-level, complex input. As P14 said, ``I thought it only do steps, but later found it does whole tasks and I will do that.''

However, participants who perceived the system as an agent would prefer to treat it as an assistant or ``you,'' expecting it to understand high-level commands, or even answer questions related to the environment. For example, participants mentioned that talking to this system is similar to ``talk[ing] normally and intuitively like you would to a person'' (P2). When there is confusion, participants may expect to get answers from the system. For example, P18 hoped the system could help him identify objects when he forgot their names: ``for example, I could say, What’s that green-striped object? and it could respond, It’s a watermelon.''

Additionally, some participants treated the virtual hand as an extension of their body that just ``behaves like my arm'' (P12), which could enhance immersion in the virtual environment. P13 said,
\begin{quote}
    \textit{``Essentially, you want to immerse your hand into this [...] virtual environment. It’s about bridging the gap between you and the virtual space, allowing direct interaction with what’s inside, like in Minecraft or other virtual worlds. You'd want to `put your hand in' and interact with objects or perform tasks as if you were physically present in that environment.''}
\end{quote}

\subsection{RQ5: What other expectations do participants have for the system?}
In the free exploration session, many participants tried commands requiring multimodal understanding of the environment, context, and how different objects are related to each other. For example, P18 wanted to say ``like the dog'' to press the heart button on top of the image viewer window. This expectation of environment understanding also extends to how the physics work in the virtual environment, such as ``throw the apple to the dog'' and ``catch the apple right before it hits the table.'' 
Commands based on visual descriptions were also used. As P19 mentioned,
\begin{quote}
    \textit{``For example, if I forgot the name of an object, like a watermelon, and describe it as a `green-striped round object,' I’d expect it to recognize the object. In real life, there are moments when you can’t recall names, so being able to describe objects and have the system understand would be helpful.''}
\end{quote}

Participants also tried additional gestures that the system did not yet directly support, such as roll, throw, flip, wipe, and detailed finger controls. While there was no direct match to these gestures, HandProxy still tried to utilize existing gestures to reproduce the command to its best. For example, the system performed a grab and rotate to ``flip a basket.''
In addition, participants also suggested other hand controls that could be useful, including bimanual control (P1, P2, P13), 
supporting more precise, affordance-dependent object interaction (P2, e.g., grab the basket handle/side), and detailed individual finger controls (P5). 
Additionally, participants would like the system to understand object states and conditions in the environment. For example, P20 tried ``if the photo has been resized, press the confirm button,'' 
This also includes monitoring the state of the object while being manipulated by the virtual hand. For example, P13 mentioned that, ``There was one time when it was trying to move the apple but didn't realize it no longer had the apple in its hand,'' if the system was aware of the hand state, it could automatically correct this error. These require understanding and monitoring the object state to relieve users from ``monitoring conditions constantly.'' (P19) 
Participants also emphasized the importance of additional feedback to help users identify and recover from errors. For example, P6 recommended underlining the parts in user's command that are confusing or uninterpretable, enabling users to rephrase their input more effectively in subsequent attempts. P1 suggested having the system predict and suggest possible next moves after the command, such as showing all afforded hand interactions once the user grabbed an object. Additionally, participants noted that improving the speech recognition system and reducing overall latency could enhance the system's usability. {\m Current limitations and possible extentions on HandProxy are further discussed in \autoref{sec:discussion}.}
\section{Discussion and Future Work}
\label{sec:discussion}Here, we expand on our key findings and discuss their implications for future work, including handling ambiguity, system transparency, supporting bimanual interaction, and creating a unified accessibility API for hand interaction. 

\subsection{Handling Ambiguity in Interaction Commands}
Ambiguity is inherently a part of the natural language, as certain commands could be interpreted in different ways. Addressing this challenge requires understanding the user’s true intentions. Based on observations from the user study, two key questions arise: \textit{(i)} what additional information can be leveraged to resolve ambiguity, and \textit{(ii)} how should we balance the system's ability to disambiguate and requesting clarifications from users. 

During the user study, the commands that were considered ``ambiguous'' by the system were often due to insufficient understanding of the environment and the target objects, including visual features, complex spatial referencing, and object affordance. {\m The current disambiguation design is mostly focusing on disambiguating duplicate objects. However, as discovered in the user study, additional ambiguity could come from the use of synonyms, or commands that have multiple interpretations. }
In cases of ambiguity, the system sometimes took initiatives and filled in the necessary information. While this worked in some cases, it inevitably caused unexpected behavior (e.g., interpreting ``turn the volume slider to maximum'' as rotate volume knob to the right, rather than moving up). This highlights a design challenge that even with additional information about the environment, it is important to determine what can and should be automatically inferred by the system, and what should be prompted to users for clarification. An avenue for future work is to investigate how the system should balance automated inference and user clarification to consider both accuracy and user effort.

\subsection{Supporting Multiple Levels of Interaction Control}
\begin{figure}[t]
    \centering
         \includegraphics[width=\linewidth]{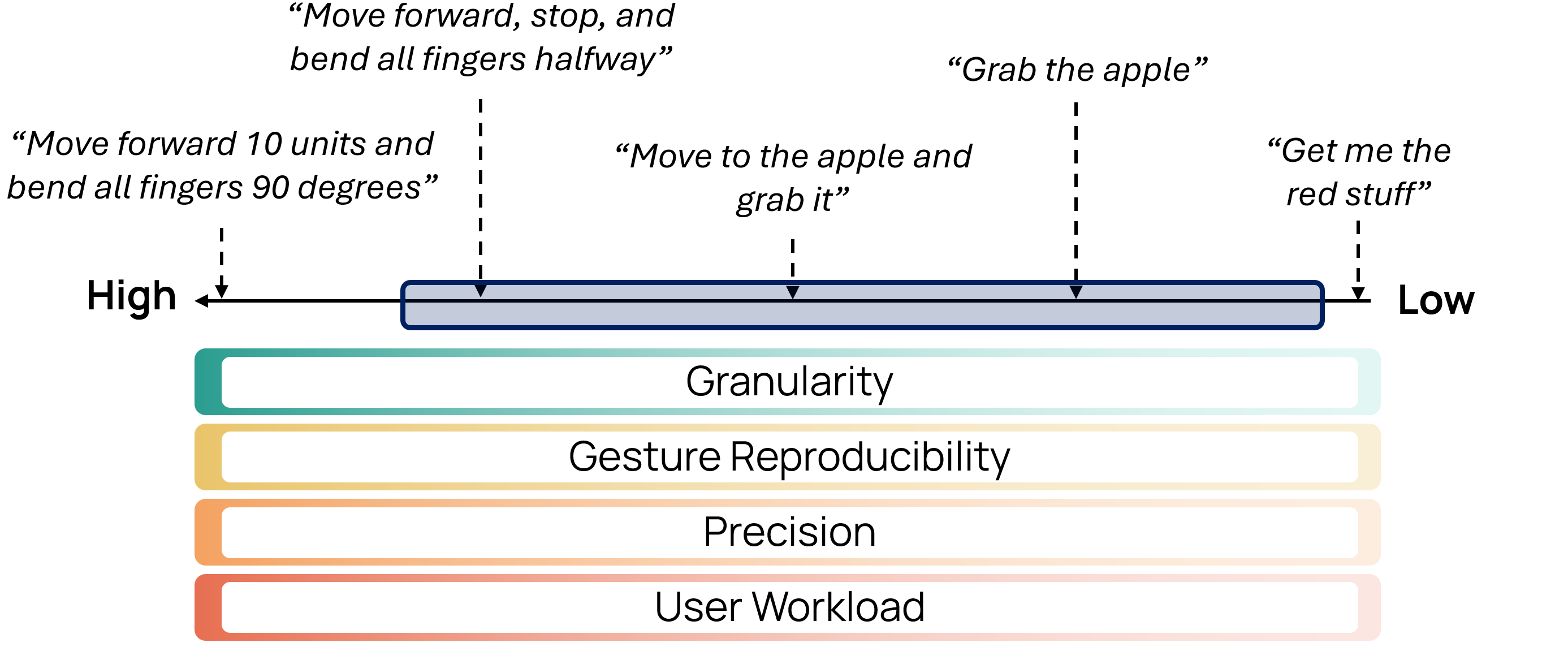}
         \vspace{-.5pc}
    \caption{{\m Spectrum of command granularity, with an estimated preferred range highlighted on the spectrum.}}
    \Description{A spectrum from high to low of command granularity, gesture reproducibility, possible ambiguity, and user workload. From high to low granularity, command examples include "move forward 10 units and bend all fingers 90 degrees, move forward stop and bend all fingers half way, move to the apple and grab it, grab the apple, and get me the red stuff. An ideal range of supported granularity should not be too specific and also not too generic, roughly 10\% to 95\% along the spectrum.}
    \label{fig:dis_spectrum}
\end{figure}

Hand interaction commands can vary in granularity along the spectrum shown in \autoref{fig:dis_spectrum}, with trade-offs in gesture reproducibility, ambiguity, user workload, and system capability. At the high-granularity end of the spectrum, users have full, precise control of the hand. This allows users to reproduce almost any hand gesture and even fine-tune or customize gestures. While such commands are less ambiguous, they impose a higher cognitive and physical workload on users due to the detailed input required. Conversely, at the low-granularity end of the spectrum, users can issue abstract, high-level interaction goals, leaving the system to determine how and what to do to achieve the goal. While this reduces user effort and simplifies command input, it comes with the trade-offs of reduced precise control, increased ambiguity, and a greater demand on the system’s interpretation capabilities.

To design effective and efficient speech to hand controls, it is important to define an appropriate range of supported commands that balances granularity and usability. We observed and estimated a preferred range for hand controls from our study, illustrated in \autoref{fig:dis_spectrum}. The preferred range skews toward the low-granularity end of the spectrum, with spacings on both left and right. The left end of the spectrum, although having more hand control possibilities, was not included because of the control complexity. Although participants preferred high-level commands, they may require significant system intelligence, and would likely introduce unnecessary ambiguity. These initial findings provide possible considerations for designing speech interfaces for hand interaction controls. However, further studies are needed to refine our understanding of the optimal level of command granularity and its impact on usability.

\subsection{Improving Transparency and Gesture Discoverability}
During the user study, participants often struggled to identify the specific part of their commands caused issues when errors occurred. We believe the feedback system could be improved by providing more details about how the system interprets commands, 
and should highlight the ambiguous parts and how the system interpreted them. These could be achieved through a better-designed text feedback of the recognized command, such as to color-code the words in the command by the level of ambiguity and highlight the parts that could not be interpreted. Without overwhelm users with excessive feedback, such improvements could enhance the transparency of the system, help users effectively refine their inputs and improve their overall experience. 

In addition to addressing errors, the interactions' discoverability remains an area for improvement. While flexible speech interfaces alleviate the challenge of knowing exactly what to say, participants may still struggle to identify the possible affordances for a given object. Although HandProxy is able to infer expected gestures from high-level commands to some extent, it could be further improved by proactively showing possible interactions. For example, P3 suggested showing a list of possible next steps after each interaction, such as options like ``throw,'' ``squeeze,'' or ``drop'' after an apple is grabbed. This could not only help users discover additional affordances but also accelerate workflows by suggesting the most relevant next steps based on the interaction history.

{\m
\subsection{Supporting Bimanual Interactions}
\label{par:dis_bimanual}
Bimanual interactions offer greater possibilities than unimanual (i.e., one-handed) interactions. Immersive apps such as Paper Birds\footnote{\url{https://www.3dar.com/p/paper-birds}} and Cubism\footnote{\url{https://www.cubism-vr.com/}} utilize both hands either collaboratively (e.g., performing a joint action) or separately (e.g., one hand for object manipulation, the other for view control). While we demonstrated the feasibility of speech-controlled one-handed interactions, the system could be extended to support bimanual interactions --- either controlling both hands or a single proxy hand collaborating with the user’s real hand.

Future work could expand the design space for bimanual interactions along two key dimensions: \textit{(i)} the type of bimanual interactions, and \textit{(ii)} the level of proxy controls. For interaction types, Yamagami et al. \cite{two_in_one} categorized bimanual interactions into symmetric in-phase (e.g., jump-roping), symmetric out-of-phase (e.g., climb a ladder), asymmetric coordinated (e.g., swing a golf club), and asymmetric uncoordinated (e.g., use two swords at the same time). Regarding the level of proxy controls, the system could be designed to perform bimanual interactions directly through two proxy hands, or employ one proxy hand that monitors, interprets, and collaborates with the user's one-handed input. This expanded design space could better support bimanual interactions.

\subsection{Limitation on Applicable Use Cases}
While the proposed approach is designed to generalize across a wide range of interaction scenarios, it may not always offer substantial advantages over alternative methods. For more direct and less hand-dependent interactions -- such as basic system operations (e.g., power on/off) or standard UI controls (e.g., closing windows) -- a traditional speech interface with direct command mapping may be more effective. Interactions requiring high precision or complexity, such as fine-grained rotations in training or simulation-based XR applications, may pose challenges when relying solely on the proposed speech interface. Furthermore, although our approach leverages commonly used gestures across various immersive applications, the current implementation does not readily accommodate applications that heavily depend on customized, non-standard hand gestures. For example, the spell-casting game Drakheir Hands of Wizard\footnote{\url{https://www.meta.com/experiences/drakheir-hands-of-wizard}} requires specialized gesture inputs that are not easily replicated within the existing system. These observations suggest that further enhancements are needed to support a broader range of use cases, particularly those involving diverse gesture types, varying levels of precision, and additional control requirements. 

As an initial exploration of the proxied interaction paradigm for speech interfaces, HandProxy shows the potential of using virtual hands to broaden interaction possibilities. Future work could address current limitations through the integration of multimodal input strategies that adaptively optimize interaction based on task type and complexity. Additional improvements might include support for runtime gesture recording to enable customizable gesture macros, as well as integration with system-level APIs to facilitate cross-application control.

\subsection{Towards an Accessible Interface for Hand Interactions}
A key motivation of this work is to find an expressive, flexible alternative to hand interactions in cases of situational impairments, ability mismatches, or user's varied preferences. While this work demonstrated how speech interfaces can be enhanced to achieve this goal, it brings up a broader question: can we define a unified control interface for possible hand interactions in the virtual environment --- just like defining an API for 2D cursor controls, so that the interactions can be mapped to a broader range of input modality setups, potentially accommodating a wide range of user's abilities and preferences?

To achieve this, it is important to create a comprehensive, well-defined set of vocabularies for hand interactions in the virtual environment. Then, the direct mapping between the hand input and an arbitrary input modality could be simplified to the mapping to this shared vocabulary. This approach could enable flexible and extensible mapping of hand input to other input setups, including multimodal configurations, that could be used to provide a more accessible way of interacting with the virtual environment. 
}
\section{Conclusion}
We presented HandProxy, a system that enables users to control a virtual proxy hand using natural speech commands, allowing it to perform various hand interactions on the user's behalf. To achieve this, we defined a set of hand control primitives and demonstrated how different hand interactions can be composed by combining these primitives. Building on this structure, we implemented HandProxy as a real-time system that supports the continuous streaming and execution of user commands with varying levels of granularity. Through a user study with 20 participants, we demonstrated that HandProxy effectively enables users to complete a wide range of tasks typically designed for direct hand interactions, and showed that HandProxy is able to interpret diverse command variations. Additionally, we explored user strategies, preferences, and expectations regarding speech-driven hand interactions. Finally, we reflected on key findings from the study and discussed their implications for future developments, including resolving ambiguity in user commands, supporting varying levels of interaction control, enhancing system transparency and gesture discoverability, supporting bimanual interactions, and directions towards an accessible interface in virtual environment. This work demonstrates the potential of speech interfaces, augmented by interaction proxies, to expand their capabilities and facilitate more expressive interactions. Our findings highlight new possibilities for initiating expressive interactions through speech interfaces and point to future directions for enhancing usability, adaptability, and intelligent interaction proxies in virtual environments.

{\m
\appendix
\section{Example Gesture Metadata File}
\label{ap:gesture_metadata}

\begin{quote}{\footnotesize\ttfamily
\{

\hspace{0.5cm}"name": "cut",

\hspace{0.5cm}"data\_format": "unified",

\hspace{0.5cm}"data\_source": "leap\_motion",

\hspace{0.5cm}"num\_hands": 1,

\hspace{0.5cm}"right\_hand\_data\_file": "cut.txt",

\hspace{0.5cm}"left\_hand\_data\_file": null,

\hspace{0.5cm}"is\_hold\_at\_peak": false,

\hspace{0.5cm}"is\_static": false,

\hspace{0.5cm}"interacting\_frame": 81,

\hspace{0.5cm}"interacting\_joint": ["pinky\_mcp"],

\hspace{0.5cm}"segments": [

        \hspace{1cm}\{
        
        \hspace{1.5cm}"name": "preparation",
        
        \hspace{1.5cm}"start\_frame": 0,
        
        \hspace{1.5cm}"end\_frame": 58
        
        \hspace{1cm}\},
        
        \hspace{1cm}\{
        
        \hspace{1.5cm}"name": "stroke",
        
        \hspace{1.5cm}"start\_frame": 58,
        
        \hspace{1.5cm}"end\_frame": 123
        
        \hspace{1cm}\},
        
        \hspace{1cm}\{
        
        \hspace{0.5cm}"name": "retraction",
        
        \hspace{0.5cm}"start\_frame": 123,
        
        \hspace{0.5cm}"end\_frame": 200
        
        \hspace{1cm}\}
        
    \hspace{0.5cm}]
    
\}
}
\end{quote}
}

\bibliographystyle{ACM-Reference-Format}
\bibliography{ref, uist_bib}

\end{document}